\documentclass[12pt,aps,prd,showpacs,amsmath,amssymb]{revtex4}
\input epsf
\usepackage{array}
\usepackage{epsfig}
\usepackage{amssymb}
\usepackage{amsmath}
\usepackage{graphicx,graphpap}
\allowdisplaybreaks

\begin{document}

\begin{titlepage}
\vskip1.5cm
\begin{center}
   {\Large \bf \boldmath Improved Light-cone QCD Sum Rule Analysis Of The Rare Decays $\Lambda_b\rightarrow\Lambda\gamma$ And $\Lambda_b\rightarrow\Lambda l^+l^-$}
    \vskip1.3cm {Long-Fei Gan, Yong-Lu Liu, Wen-Bo Chen, Ming-Qiu Huang}
  \vskip0.5cm
        {College of Science, National University of Defense Technology, Changsha, Hunan 410073, People's Republic of China} \\
\vskip1.5cm


\vskip2cm

{\large\bf Abstract\\[10pt]} \parbox[t]{\textwidth}{We present a systematic light-cone QCD sum rule study of the exclusive rare radiative decay $\Lambda_b\rightarrow\Lambda\gamma$ and rare semileptonic decay $\Lambda_b\rightarrow\Lambda l^+l^-$ within the framework of the standard model. Although some LCSR studies on these rare processes can be found in different literatures, it is necessary to reanalyze them systematically for the reason that either the baryonic distribution amplitudes are improved or different interpolating currents for the $\Lambda_b$ baryon may lead to quite different results. In addition, the rare process $\Lambda_b\rightarrow\Lambda\gamma$ has not yet been analyzed by LCSR with the Ioffe-type current. Taking all these reasons into account, we perform LCSR calculations of both the processes with two types of interpolating currents. Our calculations show that the choice of the interpolating current for the $\Lambda_b$ baryon can affect the predictions significantly, especially for the rare radiative decay process.}
\newline
\newline
\newline
\end{center}

\end{titlepage}

\setcounter{footnote}{0}
\renewcommand{\thefootnote}{\arabic{footnote}}

\newpage
\section{Introduction}
\label{sec01}
The flavor-changing neutral current(FCNC) decays of $b$-quark, such as $b\rightarrow s\gamma$ and $b\rightarrow sl^+l^-$, are important probes into the flavor sector of the electroweak theory. They are forbidden at the tree level in the standard model (SM) and induced by Glashow-Iliopoulos-Maiani (GIM) mechanism via one-loop diagrams. The contribution from the loop composed of a virtual top quark and a W boson is dominant while contributions from those involved a W boson and other quarks are strongly suppressed. Therefore, these types of processes can provide valuable information on the Cabibbo-Kobayashi-Maskawa (CKM) matrix elements $V_{ts}$ and $V_{tb}$ and possibly open a window to new physics beyond the SM.

As a matter of fact, a huge amount of work has been carried out both experimentally and theoretically in the investigations of these types of decays over the past two decades, especially on the mesonic sector \cite{LX08,Wang09}. Experimentally, the data on the rare radiative $B$ decays has been unceasingly accumulated after the first observation of the $B\to K^*\gamma$ decay in 1993 by CLEO \cite{CLEO1}. More and more precise measurements of the exclusive and inclusive branching fractions have also been reported \cite{CLEO,BABAR,BELLE}. Theoretically, various approaches have been employed to investigate the inclusive
$b\rightarrow s\gamma$ and the exclusive $B\to K^*\gamma$ processes within and beyond the SM (for a review see Ref. \cite{hurth} and references therein). Predictions for the inclusive decay rates within the SM are in good agreement with experimental data. But for the exclusive rare $B$ decay rates, the predictions need to be improved for that the nonperturbative contributions are deeply involved and we can only calculate them in some model-dependent ways.

In contrast to those of $B$ mesons, the rare decay processes of b-baryons have attracted much less attention for the relatively complicated quark dynamics, as well as the insufficiency of experimental data. But they can help to extract more experimental information about the heavy beauty quark inside the hadron, such as the hadronic spin polarization during hadronization and the helicity structure of the couplings at the quark level, which is impossible for the mesonic decays. In point of fact, the rare decays of b-baryons can provide a new test ground for theoretical methods on the b-quark hadronization. Studies for the exclusive rare radiative $\Lambda_b$ decay can be found in the literatures using various approaches, including quark models \cite{Mannel,Mohanta,HaiYang,Chun}, QCD sum rules \cite{hcs} and the perturbative QCD method \cite{lxq}. Nevertheless, the existing theoretical predictions vary from each other and can differ even by orders of magnitude. The exclusive rare processes $\Lambda_b\rightarrow\Lambda\gamma$ and $\Lambda_b\rightarrow
\Lambda l^+l^-$ were examined with the QCD light-cone sum rule (LCSR) method in Ref. \cite{wym}. The interpolating current to the $\Lambda_b$ baryon used there is the CZ-type current. However, the distribution amplitudes of the $\Lambda$ baryon used in the numerical analysis have been improved thereafter, which may bring in a rather large effect. This has been confirmed for the calculation of electromagnetic form factors, as has been illustrated in Ref. \cite{Lenz}. The processes $\Lambda_b\rightarrow
\Lambda l^+l^-$ ($l = e,\, \mu,\, \tau$) have recently been studied by LCSR in Ref. \cite{Aliev}, using correct distribution amplitudes given in Ref. \cite{Lambdadas}. But the interpolating current they used is of a different form, compared with CZ-type and Ioffe-type currents.  As we know, the choice of the interpolating currents may affect the results to some extent \cite{lambdac,xiioffe}. Taking all these into account, we calculate the decay rate and branching ratio of the process $\Lambda_b\rightarrow\Lambda\gamma$ with both the CZ-type and the Ioffe-type currents for the $\Lambda_b$ baryon. We also reanalyze the $\Lambda_b\rightarrow
\Lambda l^+l^-$ ($l = e,\, \mu$) processes with the improved distribution amplitudes and draw a comparison between results in the cases where either the CZ-type current or the Ioffe-type current is employed in order to give more reliable predictions.

The remainder of this paper is organized as follows: Sec. \ref{sec1} is a brief introduction to the theoretical framework for our investigations. The formulas of the decay widths in terms of form factors are also derived in this section. In Sec. \ref{sec2}, we derive the light-cone sum rules for the relevant form factors, where both the CZ-type and the Ioffe-type interpolating currents for the $\Lambda_{b}$ baryon are considered.  Sec. \ref{sec3} is devoted to the numerical analysis and discussions, a summary is then attached at the end of this section.
\section{Theoretical framework}\label{sec1}
The effective Hamiltonian for the $b\rightarrow s\gamma$ transition in the standard model at scales $\mu = O (m_{b})$ is given by \cite{hamiltonian}
\begin{equation}\label{Lagrangian01}
H_{\mathrm{eff}}(b\rightarrow s\gamma) = -4\frac{G_F}{\sqrt{2}}  V_{t s}^\ast  V_{tb}
  C_{7}(\mu)  {\cal O}_7(\mu),
\end{equation}
with the operator
\begin{equation}
{\cal O}_{7} = \frac{e }{ 16\ \pi^2}\bar{s}\ \sigma_{\mu \nu}\
(m_{b} R+ m_{s} L)\ b\ F^{\mu \nu},
\end{equation}
where $L/R=(1\mp\gamma_5)/2$ and $F^{\mu \nu}$ is the field strength tensor of the photon. $G_F$ is the Fermi coupling constant and $C_{7}(\mu)$ is the Wilson coefficient at the scale $\mu$. In order to allow for contributions from non-standard model couplings, the operator ${\cal O}_{7}$ can be written in a more general form as
\begin{equation}
{\cal O}_{7} = \frac{e}{32\ \pi^2}m_{b}\bar{s}\ \sigma_{\mu \nu}\ (
g_V + \gamma_5g_A)\ b\ F^{\mu \nu},
\end{equation}
where $g_V=1+m_s/m_b$ and $g_A=1-m_s/m_b$ in the SM.
Similarly, the effective Hamiltonian relative to the process $b\rightarrow s l^+l^-$ can be expressed as
\begin{eqnarray}\label{Lagrangian02}
H_{\mathrm{eff}}(b\rightarrow s l^+l^-)&&=\frac{G_F V_{tb}V_{ts}^*\alpha_{em}}{2\sqrt 2\pi}\Big\{-C_7^{\mathrm{eff}}\frac{2}{q^2}\bar si\sigma_{\mu\nu}q^\nu(m_bR+m_sL)b\bar l\gamma^\mu l\nonumber\\
&&+C_9^{\mathrm{eff}}\bar s \gamma_\mu Lb\bar l\gamma^\mu l+C_{10}^{\mathrm{eff}}\bar s \gamma_\mu Lb\bar l\gamma^\mu\gamma_5 l\Big\}.
\end{eqnarray}
Here we have neglected terms proportional to $V_{ub} V_{us}^*$ in the effective Hamiltonian since the ratio $\displaystyle \left |{V_{ub}V_{us}^* \over V_{tb}V_{ts}^*}\right |$ is of the order $10^{-2}$.
As well known, the amplitudes of the rare processes $\Lambda_b\rightarrow\Lambda\gamma$ and $\Lambda_b\rightarrow\Lambda l^+l^-$ are determined by the matrix elements of the effective Hamiltonians sandwiched between the initial and final states at the hadron level, which can not be calculated straightforwardly from first principles for the nonperturbative property of QCD. They can generally be parameterized in terms of form factors as follows:
\begin{eqnarray}\label{matrix01}
&&\langle\Lambda_b(P')|i\bar b\sigma_{\mu\nu}q^\mu s|\Lambda(P)\rangle=\bar \Lambda_b(P')(f_1\gamma_\nu+f_2 i\sigma_{\mu\nu}q^\mu+f_3q_\nu)\Lambda(P),\nonumber\\
&&\langle\Lambda_b(P')|i\bar b\sigma_{\mu\nu}q^\mu\gamma_5 s|\Lambda(P)\rangle=\bar \Lambda_b(P')(g_1\gamma_\nu+g_2 i\sigma_{\mu\nu}q^\mu+g_3q_\nu)\gamma_5\Lambda(P),\nonumber\\
&&\langle\Lambda_b(P')|\bar b\gamma_\mu s|\Lambda(P)\rangle=\bar \Lambda_b(P')(F_1\gamma_\nu+F_2 i\sigma_{\mu\nu}q^\mu+F_3q_\nu)\Lambda(P),\nonumber\\
&&\langle\Lambda_b(P')|\bar b\gamma_\mu\gamma_5 s|\Lambda(P)\rangle=\bar \Lambda_b(P')(G_1\gamma_\nu+G_2 i\sigma_{\mu\nu}q^\mu+G_3q_\nu)\gamma_5\Lambda(P),
\end{eqnarray}
where $\Lambda_b$ and $\Lambda$ on the right-hand side are the spinors of the $\Lambda_b$ baryon and the $\Lambda$ baryon, respectively. $P$ is the momentum of the $\Lambda$ baryon, $P'=P+q$ is the momentum of the $\Lambda_b$ baryon, and $q$ is the momentum transfer.

The decay mode $\Lambda_b\rightarrow \Lambda\gamma$ takes place with a real photon in its final state, which makes the parametrization of the hadronic part a little simpler. The hadronic matrix element of the relevant operator can now be parameterized in the following covariant form:
\begin{eqnarray}\label{matrix02}
\langle\Lambda_b(P,s)|\bar b\;\sigma_{\mu\nu}q^\nu ( g_V + \gamma_5
g_A ) s|\Lambda(P^\prime,s^\prime \rangle&=& \bar
\Lambda_b(P,s)\sigma_{\mu\nu}q^\nu \left(g_V f_2+\gamma_5 g_A
g_2\right)\Lambda(P^\prime,s^\prime).
\end{eqnarray}
Utilizing this parameterized hadronic element, the formula for the decay rate of the process $\Lambda_b\rightarrow \Lambda\gamma$ can be immediately derived in terms of the two form factors $f_2$ and $g_2$ as below:
\begin{equation}\label{decayrate01}
\Gamma(\Lambda_b\rightarrow\Lambda\gamma)=\frac{G_F^2|V_{tb}V_{ts}^*|^2
\alpha_{em}|C_7|^2 m_b^2}{32\pi^4}\left(
\frac{M_{\Lambda_b}^2-M_\Lambda^2}{M_{\Lambda_b}} \right )^3(g_V^2
f_2^2 + g_A^2 g_2^2).
\end{equation}
Note that we have not considered the long-distance contribution to the processes coming from $c\bar c$ resonances, which results mainly in the correction to the Wilson coefficient $C_7$. These resonant (long-distance) contributions were estimated using the vector meson dominance model in Ref. \cite{Deshpande}, and it has been shown that they are suppressed roughly by a factor of $10$ due to the emission of a real photon in the $b\rightarrow s\gamma$ process.

Similarly, the differential decay width of the process $\Lambda_b\rightarrow \Lambda l^+l^-$ at momentum transfer $q^2$ can also be expressed with the form factors defined above in Eq. (\ref{matrix01}) as
\begin{eqnarray}\label{decayrate02}
\frac{d \Gamma (\Lambda_b\rightarrow \Lambda l^+l^-)}{d q^2} & = & \frac{(G_F|V_{tb}||V_{ts}|\alpha_{em})^2}{3\times 2^{10}\pi^5 M_{\Lambda_b}}\big[(\frac{q^2}{M_{\Lambda_b}^2}-(1+r^2))^2-4 r^2\big]^{1/2}
\nonumber\\ & & \bigg\{-\frac{4 C_7 C^*_7}{q^4} \big\{(q^2-M_{\Lambda_b}^2 (r+1)^2) \big[(q^2+2 M_{\Lambda_b}^2 (r-1)^2) g_2^2 q^2 \nonumber\\ & & +6 M_{\Lambda_b} (r-1) g_1 g_2 q^2+(2 q^2+M_{\Lambda_b}^2 (r-1)^2) g_1^2\big] (m_b-m_s) ^2 \nonumber\\ & & +\big[-(r^2-1)^2 M_{\Lambda_b}^4-q^2 (r^2-6 r+1) M_{\Lambda_b}^2+2 q^4\big] f_1^2 (m_b+m_s) ^2 \nonumber\\ & & +q^2 \big[-2 (r^2-1)^2 M_{\Lambda_b}^4+q^2 (r^2+6 r+1) M_{\Lambda_b}^2+q^4\big] f_2^2 (m_b+m_s) ^2 \nonumber\\ & & -6 M_{\Lambda_b} q^2 (q^2-M_{\Lambda_b}^2 (r-1)^2) (r+1) f_1 f_2 (m_b+m_s)^2 \big\}\nonumber\\ & &
+\frac{2}{q^2}(C_9 C^* _7+C_7 C^* _9)\big\{(q^2-M_{\Lambda_b}^2 (r-1)^2) \big[(q^2+2 M_{\Lambda_b}^2 (r+1)^2)f_2 F_2 \nonumber\\ & & -3 M_{\Lambda_b} (r+1)f_2 F_1\big] (m_b+m_s) q^2+(q^2-M_{\Lambda_b}^2 (r+1)^2) \nonumber\\ & & \big[ 3 M_{\Lambda_b} (r-1) q^2(g_2 G_1+ g_1 G_2)+(q^2+2 M_{\Lambda_b}^2 (r-1)^2)g_2 G_2 q^2 \nonumber\\ & & +(2 q^2+M_{\Lambda_b}^2 (r-1)^2)g_1 G_1\big] (m_b-m_s) +(q^2-M_{\Lambda_b}^2 (r-1)^2) \nonumber\\ & & \big[(2 q^2+M_{\Lambda_b}^2 (r+1)^2)f_1 F_1-3 M_{\Lambda_b} q^2 (r+1)f_1 F_2\big] (m_b+m_s)\big\} \nonumber\\ & &
-(C_9 C^*_9+C_{10} C^*_{10})\big\{\big[-2 (r^2-1)^2 M_{\Lambda_b}^4+q^2 (r^2+6 r+1) M_{\Lambda_b}^2+q^4\big] F_2^2 q^2 \nonumber\\ & & -6 M_{\Lambda_b} (q^2-M_{\Lambda_b}^2 (r-1)^2) (r+1) F_1 F_2 q^2+\big[-(r^2-1)^2 M_{\Lambda_b}^4 \nonumber\\ & & -q^2 (r^2-6 r+1) M_{\Lambda_b}^2+2 q^4\big] F_1^2+(q^2-M_{\Lambda_b}^2 (r+1)^2) \nonumber\\ & & \big[(q^2+2 M_{\Lambda_b}^2 (r-1)^2) G_2^2 q^2+6 M_{\Lambda_b} (r-1) G_1 G_2 q^2 \nonumber\\ & & +(2 q^2+M_{\Lambda_b}^2 (r-1)^2) G_1^2\big]\big\}\bigg\},
\end{eqnarray}
where $r=M_{\Lambda}/M_{\Lambda_b}$ is the ratio between masses of the $\Lambda$ and $\Lambda_b$ baryons. We should note here that the form factors $F_3$ and $G_3$ vanish due to the conservation of the vector current $\bar l\gamma_\mu l$ so that they are absent in Eqs. (\ref{decayrate01}) and (\ref{decayrate02}). In order to calculate these decay rates, we have to firstly estimate all the form factors appearing in these equations. They are nonperturbative quantities which should be estimated with nonperturbative approaches. In the next section, we will employ the light-cone sum rule approach to estimate them.

\section{ Light-cone QCD Sum Rules for the form factors}
\label{sec2}
It has been shown in Refs. \cite{IIG} that there is some randomness in the choice of interpolating currents for baryons with definite quantum numbers in QCD sum rule calculations, and the practical criterion is that the coupling between the interpolating current and the given state must be strong enough. In the following LCSR analysis, we will adopt two types of currents to interpolate the $\Lambda_b$ baryon state: the CZ-type current \cite{chernyak}
\begin{eqnarray}\label{current01}
{j_\Lambda}(x)&=&\epsilon_{ijk}(u^iC\gamma_5\rlap/zd^j)\rlap/ zb^k,
\end{eqnarray}
and the Ioffe-type current \cite{Ioffe}
\begin{eqnarray}\label{current02}
{\tilde j_\Lambda}(x)&=&\epsilon_{ijk}(u^iC\gamma_5\gamma_\mu d^j)\gamma_\mu b^k,
\end{eqnarray}
where $C$ denotes the charge conjugation matrix, and $i,j,k$ are the color indices. The auxiliary four-vector $z$, which satisfies $z^2=0$, is introduced to project the main contribution out onto the light-cone. The coupling constants of the interpolating currents between the $\Lambda_b$ baryon state and the vacuum are defined as
\begin{equation}\label{constant01}
\langle0|j_\Lambda|\Lambda_b(P',s)\rangle=f_{\Lambda_b} z\cdot
P\rlap/ z \Lambda_b(P'),
\end{equation}
and
\begin{equation}\label{constant02}
\langle0|\tilde j_\Lambda|\Lambda_b(P',s)\rangle=\lambda_{1b}M_{\Lambda_b}\Lambda_b(P'),
\end{equation}
where $\Lambda_b(P',s)$ is the $\Lambda_b$ spinor with momentum $P'$ and spin $s$. With these definitions at hand, we can now proceed to derive the QCD light-cone sum rules for the form factors mentioned above.

Let us first consider the case in which the CZ-type current is adopted. According to the general philosophy of the LCSR approach, we study the analytical property of the correlation function
\begin{equation}\label{corr}
T_\mu=i \int d^4x e^{iqx}\langle 0|{j_{\Lambda}(0)j^{\dag}_\mu(x)}|\Lambda(P,s)\rangle,
\end{equation}
at momentum transfer $q^2$ with Euclidean $m_b^2-{P^\prime}^2$ of several $\mbox{GeV}^2$. The current $j_{\Lambda}$ is the CZ-type current given above in Eq. (\ref{current01}) and  $j_\mu$ denotes the operators of the hadronic parts in the effective Hamiltonians (\ref{Lagrangian01}) and (\ref{Lagrangian02}).

Take $j_\mu=\bar s_\alpha\sigma_{\mu\nu}Rq^\nu b_\alpha$ for example, we know from Sec. \ref{sec1} that, when the process $\Lambda_b\rightarrow \Lambda\gamma$ is considered, its hadronic matrix element between the $\Lambda_b$ and $\Lambda$ states can be parameterized in terms of form factors as
\begin{eqnarray}\label{matrix03}
\langle\Lambda_b(P^\prime,s^\prime)|(\bar s_\alpha\sigma_{\mu\nu}Rq^\nu b_\alpha)^\dag|\Lambda(P,s)\rangle=\frac{1}{2}\bar{
\Lambda}_b(P^\prime,s^\prime) \sigma_{\mu\nu}q^\nu (f_2-g_2 \gamma_5)
\Lambda (P,s).
\end{eqnarray}
Inserting a complete set of intermediate states into the correlation function (\ref{corr}) and taking into account Eqs. (\ref{constant01}) and (\ref{matrix03}), we find its hadronic representation as follows:
\begin{equation}\label{phenomenum01}
z^\mu T_\mu=i\frac{f_{\Lambda_b}}{M_{\Lambda_b}^2-P^{\prime
2}}(z\cdot P^\prime)^2(f_2-g_2\gamma_5)\rlap/ z\rlap/
q\;\Lambda(P,s)+...,
\end{equation}
where $P^\prime=P-q$, and the dots denote contributions from higher resonance and continuum states. The correlation function is contracted with the light-cone vector $z^\mu$ in order to simplify the Lorentz structure and remove the terms proportional to $z^\mu$ which give subdominant contributions on the light-cone \cite{Braun,nucleon1}.

On the other hand, the correlation function can be expanded on the light-cone. To the leading order of $\alpha_s$, we calculate the correlation function straightforwardly by contracting the b quarks in Eq. (\ref{corr}). A simple derivation leads to
\begin{eqnarray}
z^\mu T_\mu&=&-\frac{1}{2}\int
d^4x\frac{d^4k}{(2\pi)^4}e^{i(q+k)\cdot x}\frac{z^\mu
q^\nu}{k^2-m_b^2}\Big\{m_b(C \gamma_5\rlap/ z)_{\alpha\beta}(\rlap/
z\sigma_{\mu\nu}(1-\gamma_5))_{\delta\omega}\nonumber\\
&& +(C \gamma_5\not z)_{\alpha\beta}(\rlap/z\rlap/
k\sigma_{\mu\nu}(1-\gamma_5))_{\delta\omega}\Big\}\langle
0|\varepsilon_{ijk}u_\alpha^i(0)d_\beta^j(0)s_\omega^k(x)|\Lambda(P,s)\rangle.
\label{zt}
\end{eqnarray}
The nonperturbative effects have been encoded in the matrix elements of the non-local operators sandwiched between the vacuum and the $\Lambda$ baryon state, which are usually parameterized by distribution amplitudes, as have been given in Ref. \cite{Lambdadas}.

Following the standard procedure of the QCD sum rule method, we match the hadronic representation and the light-cone QCD expansion series of the correlation function. With the assumption of quark-hadron duality, contributions from higher resonance and continuum states are approximated by the same dispersion integral above some effective threshold $s_0$ and can be canceled out on both sides. Then the Borel transformation in $P'^2$ is performed on both sides of the sum rules to suppress contributions of higher resonances. Finally we obtain the light-cone sum rules for the form factors $f_1$ and $f_2$ (see Appendix \ref{appendix01}). Similarly, the light-cone sum rules for the form factors $F_1$ and $F_2$ can be derived by repeating the procedure above with the operator $j_\mu=\bar s_\alpha\sigma_{\mu\nu}Rq^\nu b_\alpha$ replaced by $j_\mu=\bar s_\alpha\gamma_{\mu} b_\alpha$. Other form factors, such as $g_1$, $g_2$, $G_1$, and $G_2$, can be easily read from the relationships $g_1=-f_1$, $g_2=f_2$, $G_1=-F_1$, and $G_2=F_2$ in this case.

In the later case where the Ioffe-type current is adopted, we just need to reanalysis the correlation function (\ref{corr}) with $j_{\Lambda}$ replaced by the Ioffe-type current $\tilde j_\Lambda$ defined in (\ref{current02}). After inserting a complete set of intermediate states and using the definitions (\ref{matrix01}) and (\ref{constant02}), the hadronic representation of the correlation function now becomes
\begin{eqnarray}
z_\nu T^\nu=&&\frac{\lambda_1M_{\Lambda_b}}{M_{\Lambda_b}^2-P'^2}\{2f_1P\cdot z+2f_2P\cdot z\rlap/q_\bot +(f_1(M_{\Lambda_b}-M)-f_2q^2)\rlap/ z\nonumber\\&&-(f_1-(M_{\Lambda_b}+M)f_2)\rlap/z\rlap/q-2g_1P\cdot z\gamma_5-2g_2P\cdot z\rlap/q_\bot\gamma_5\nonumber\\
&& +(g_1(M_{\Lambda_b}-M)+g_2q^2)\rlap/z\gamma_5+(g_1+(M_{\Lambda_b}-M)g_2)\rlap/z\rlap/q\gamma_5\}\Lambda(P)+...,
\end{eqnarray}
where ``..." also stands for the contributions from higher resonance and continuum states. Here we have still taken the operator $j_\mu=\bar s_\alpha\sigma_{\mu\nu}Rq^\nu b_\alpha$ as an example. After some tedious derivations, we reach the light-cone sum rules of the form factors.

All the final sum rules for the form factors derived in both cases mentioned above are collected in Appendix \ref{appendix01}.

\section{Numerical analysis and discussions}
\label{sec3}
In order to perform the numerical calculations, we need firstly to specify the input parameters appearing in the sum rules. The coupling constants related to the distribution amplitudes of the $\Lambda$ baryon are given in Ref. \cite{Lambdadas}, which turn out to be $f_\Lambda=6.0\times10^{-3}\;\mbox{GeV}^{2}$ and $\lambda_1=0.01\;\mbox{GeV}^{2}$. As to the couplings $f_{\Lambda_b}$ and $\lambda_{1b}$, we recur to the QCD sum rules in Ref \cite{lambdac} with the replacement $m_c\rightarrow m_b$, from which we can get the values: $f_{\Lambda_b}=(3.86\pm0.12)\times10^{-3}\;\mbox{GeV}^{2}$ and $\lambda_{1b}=(0.027\pm0.03)\,\mbox{GeV}^2$.
For the mass of the $\Lambda$ and $\Lambda_b$ baryons, we use their physical values $M_\Lambda=1.116\;\mbox{GeV}$ and $M_{\Lambda_b}=5.620\;\mbox{GeV}$ \cite{PDG10}. The mass of $b$ quark we use in this paper is $m_b=4.7\;\mbox{GeV}$.

\begin{figure}
    \centering
  \begin{minipage}{7cm}
  \epsfxsize=7cm \centerline{\epsffile{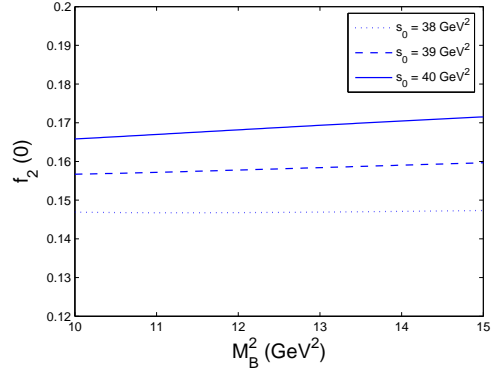}}
  \end{minipage}
  \caption{Dependence of the form factors $f_2(0)$ on the Borel parameter at $q^2 = 0\; \mbox{GeV}^2$}
  \label{fig01}
\end{figure}

With all these parameters at hand, one can proceed to compute the numerical values of the form factors. An important step in the numerical analysis of the QCD sum rules is to determine the continuum threshold $s_0$ and the Borel mass parameter $M_B^2$. The continuum threshold $s_0$ can be chosen by demanding that the continuum contribution is subdominant in comparison with that of the ground state which we are concerned about. Simultaneously, the resulting form factors should not vary drastically along with the threshold. Thus $s_0$ is generally connected with the first excited state which has the same quantum numbers as the particle we are caring about. Here we fix the threshold $s_0$ in the region $38\,\mbox{GeV}^2\le s_0\le 40\,\mbox{GeV}^2$. As for the Borel parameter $M_B^2$, we know from the LCSR that the higher twist contributions are proportional to powers like $1/(M_B^2)^n$ with $n\geq 1$. So the Borel parameter $M_B^2$ should be large enough to suppress contributions from higher twist terms. On the other hand, $M_B^2$ should not be too large, otherwise the higher resonances and continuum contributions will become dominant on the hadronic representation side. We find the results are acceptably stable in the range $10 \;\mbox{GeV}^{2}\leq {M_B}^2\leq 15\;\mbox{GeV}^{2}$, which is just the working window. As an example, we plot the dependence of the form factor $f_2(0)$ on the Borel parameter in Fig. \ref{fig01}.

\begin{figure}
\begin{minipage}{7cm}
\epsfxsize=7cm \centerline{\epsffile{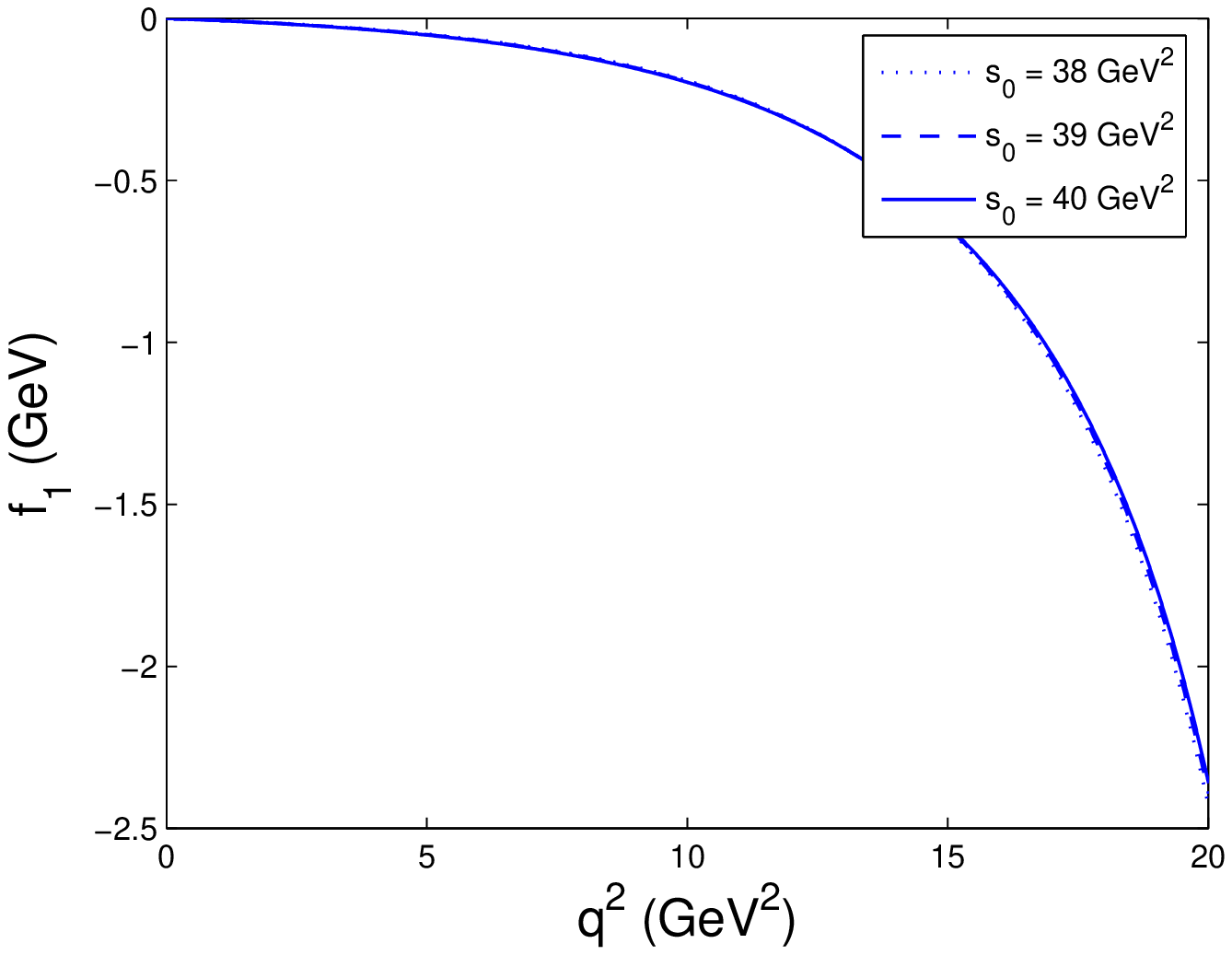}}
\end{minipage}
\begin{minipage}{7cm}
\epsfxsize=7cm \centerline{\epsffile{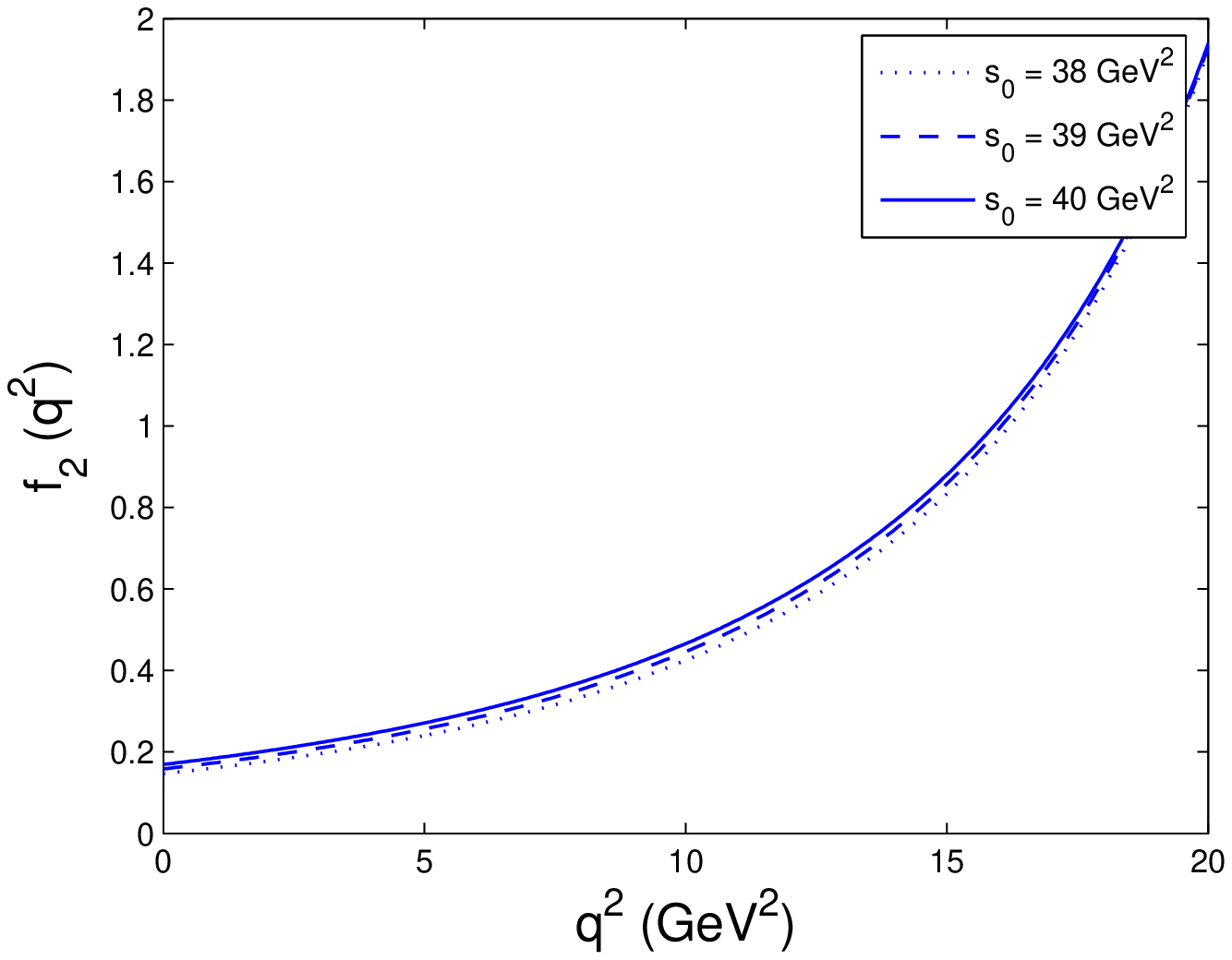}}
\end{minipage}
\begin{minipage}{7cm}
\epsfxsize=7cm \centerline{\epsffile{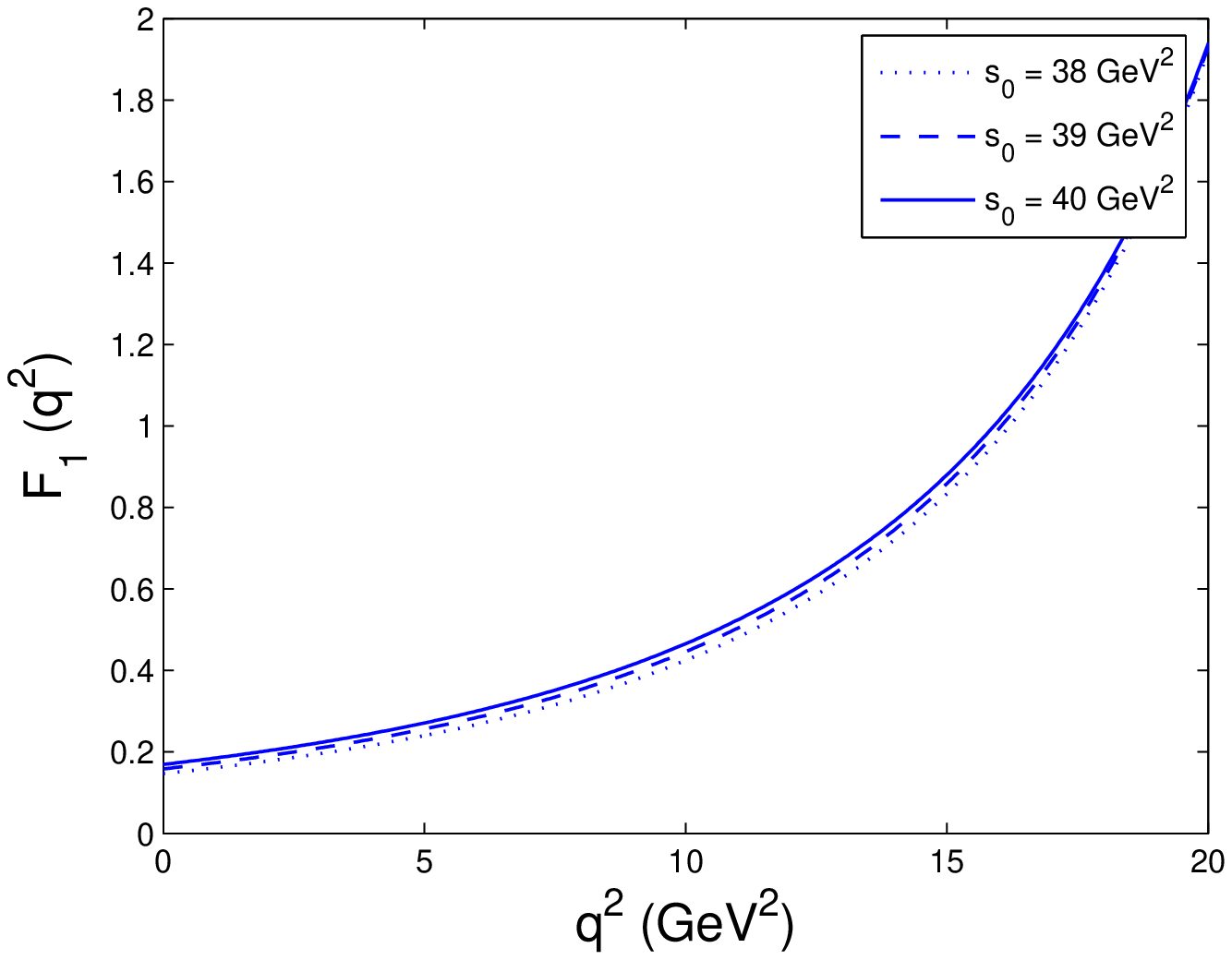}}
\end{minipage}
\begin{minipage}{7cm}
\epsfxsize=7cm \centerline{\epsffile{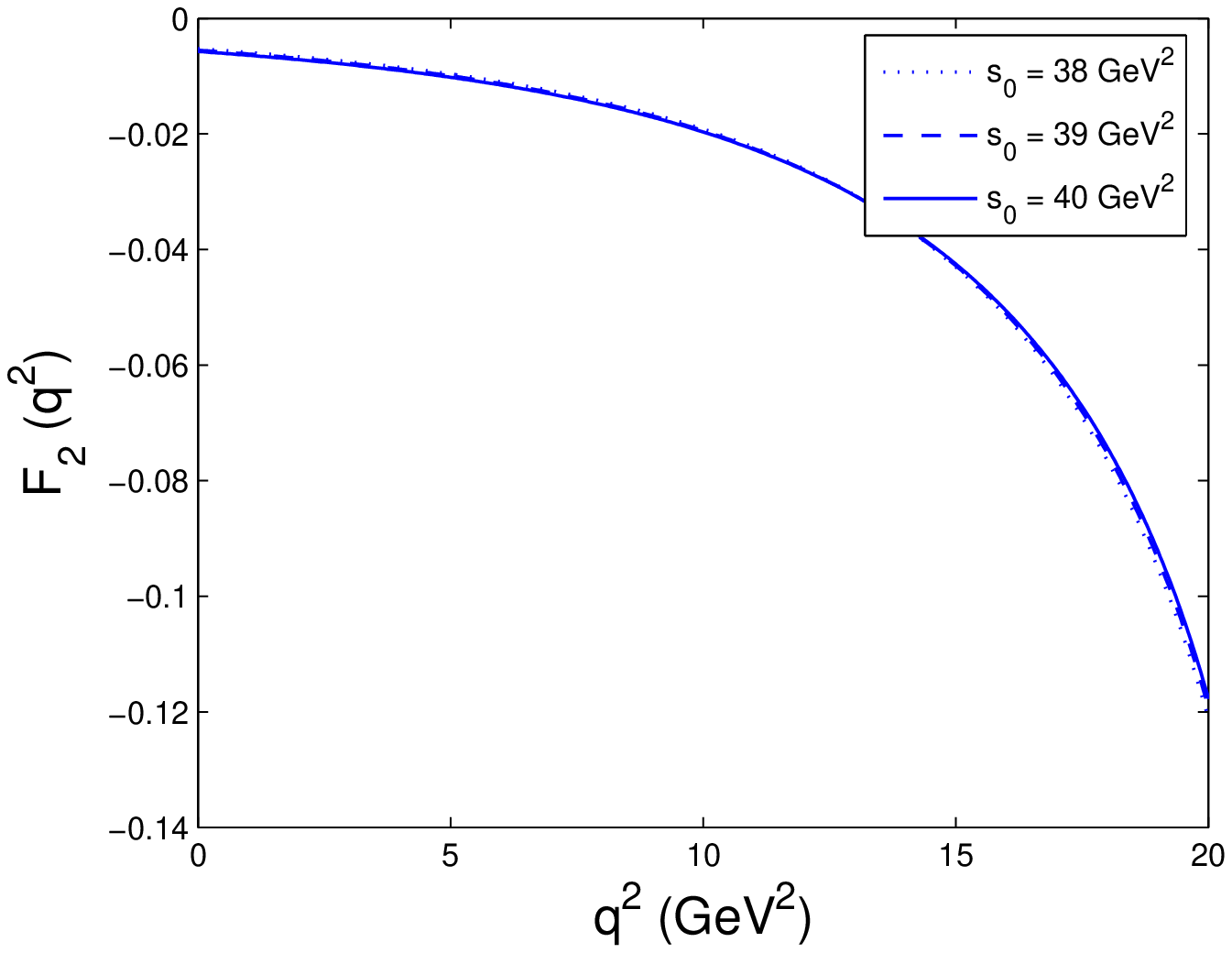}}
\end{minipage}
\caption{\quad Predictions of the form factors on the momentum transfer $q^2$ with $s_0=38,\,39,\,40\,\mbox{GeV}^2$ when CZ-type current is used. }\label{fig02}
\end{figure}

\begin{figure}
\begin{minipage}{7cm}
\epsfxsize=7cm \centerline{\epsffile{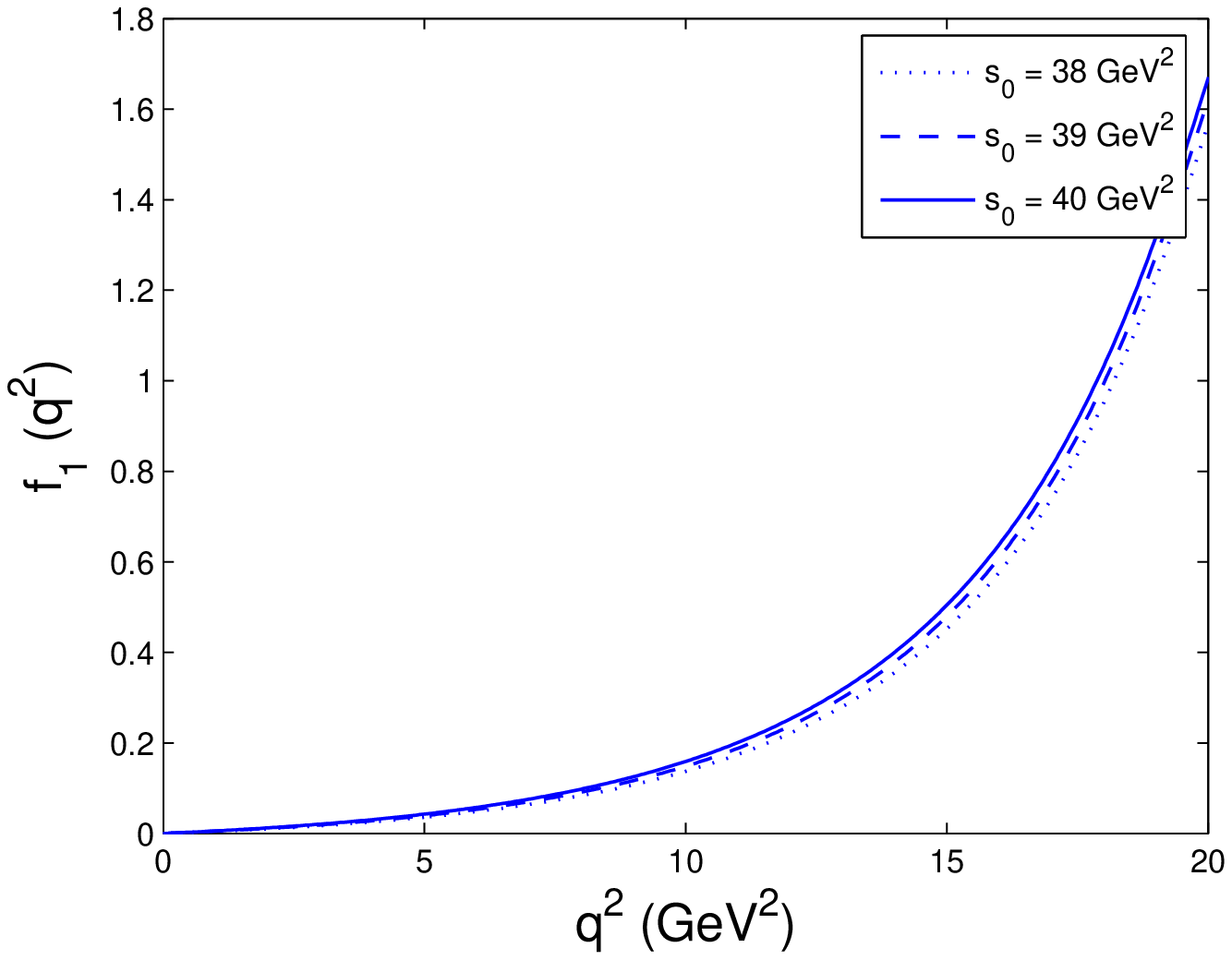}}
\end{minipage}
\begin{minipage}{7cm}
\epsfxsize=7cm \centerline{\epsffile{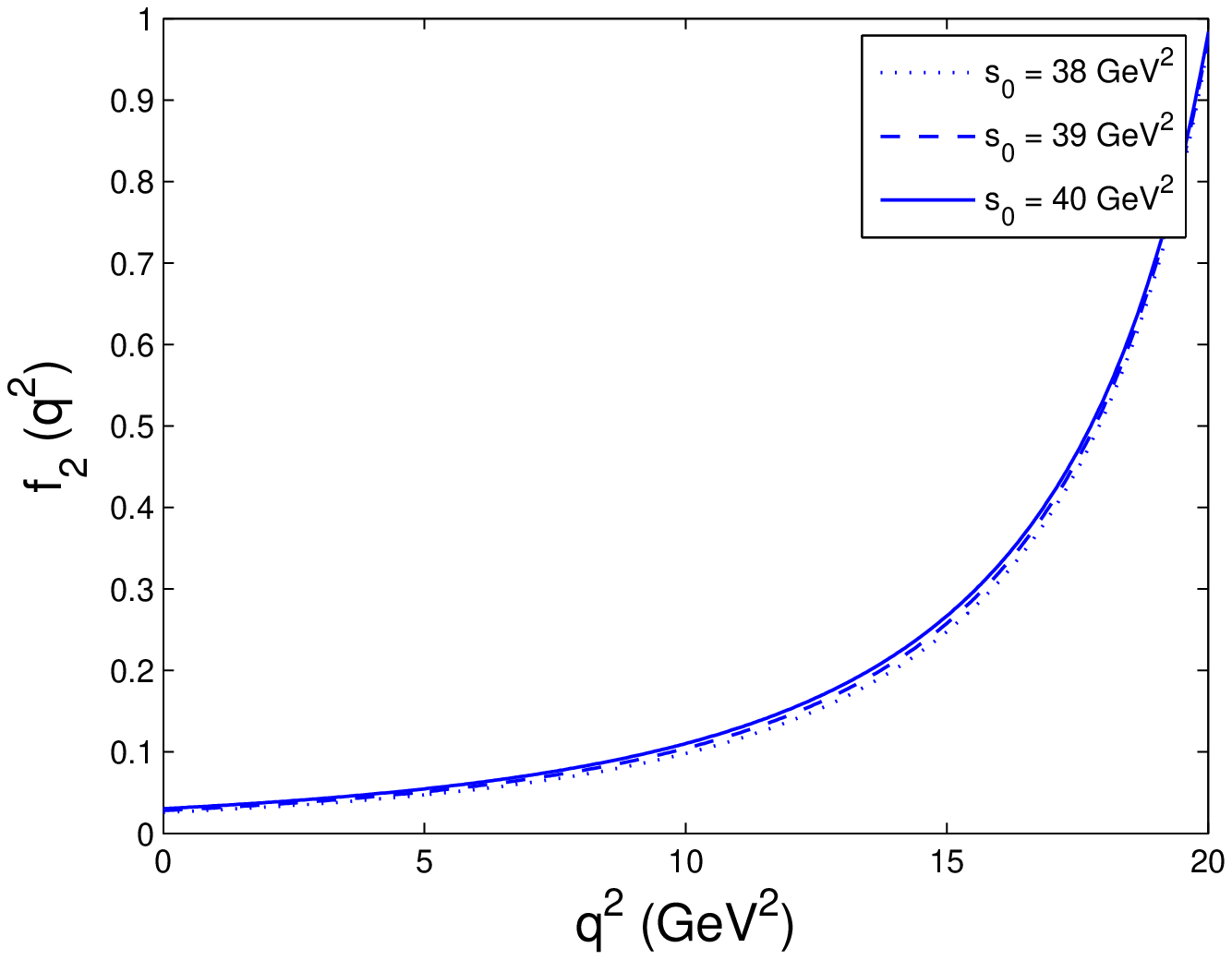}}
\end{minipage}
\begin{minipage}{7cm}
\epsfxsize=7cm \centerline{\epsffile{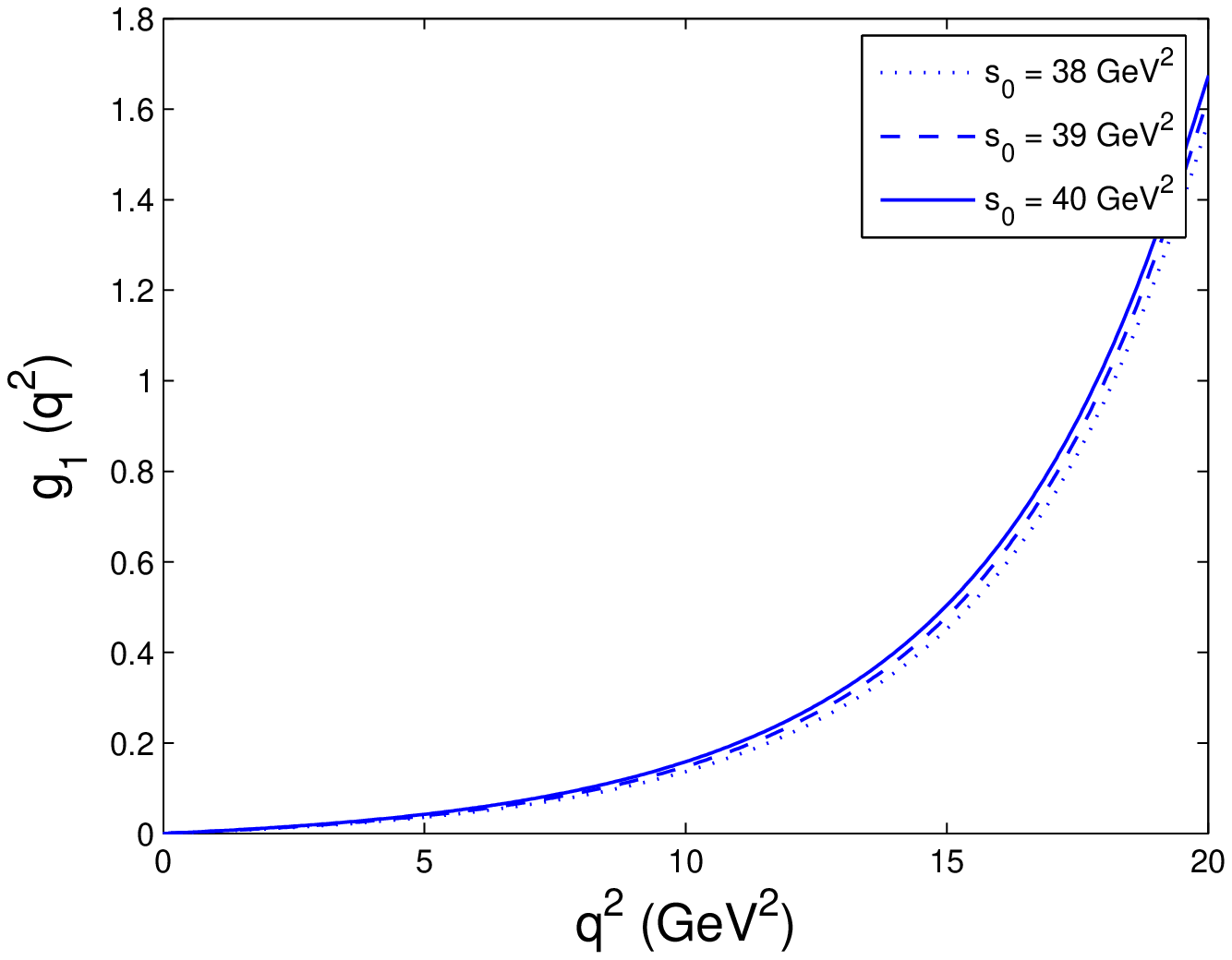}}
\end{minipage}
\begin{minipage}{7cm}
\epsfxsize=7cm \centerline{\epsffile{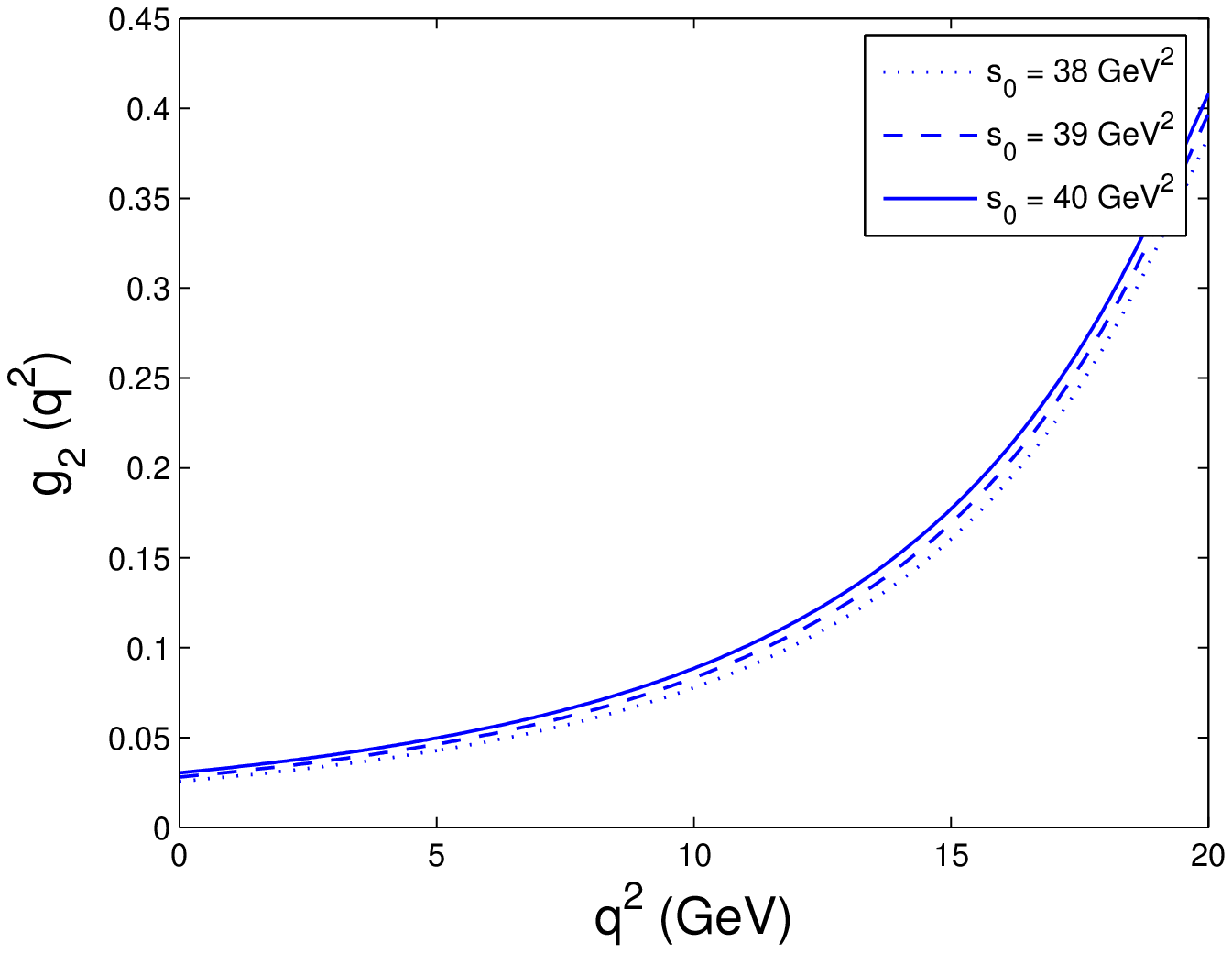}}
\end{minipage}
\begin{minipage}{7cm}
\epsfxsize=7cm \centerline{\epsffile{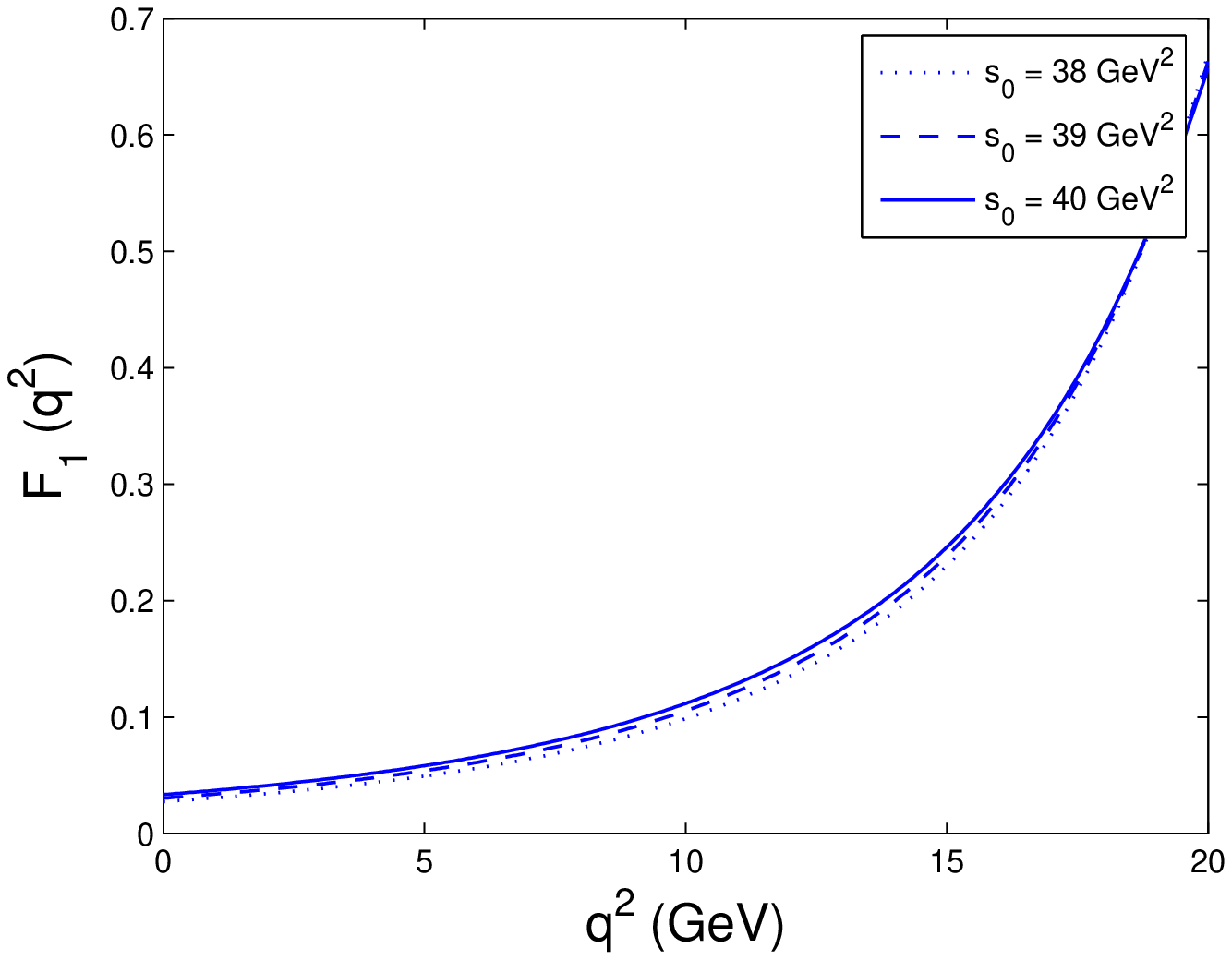}}
\end{minipage}
\begin{minipage}{7cm}
\epsfxsize=7cm \centerline{\epsffile{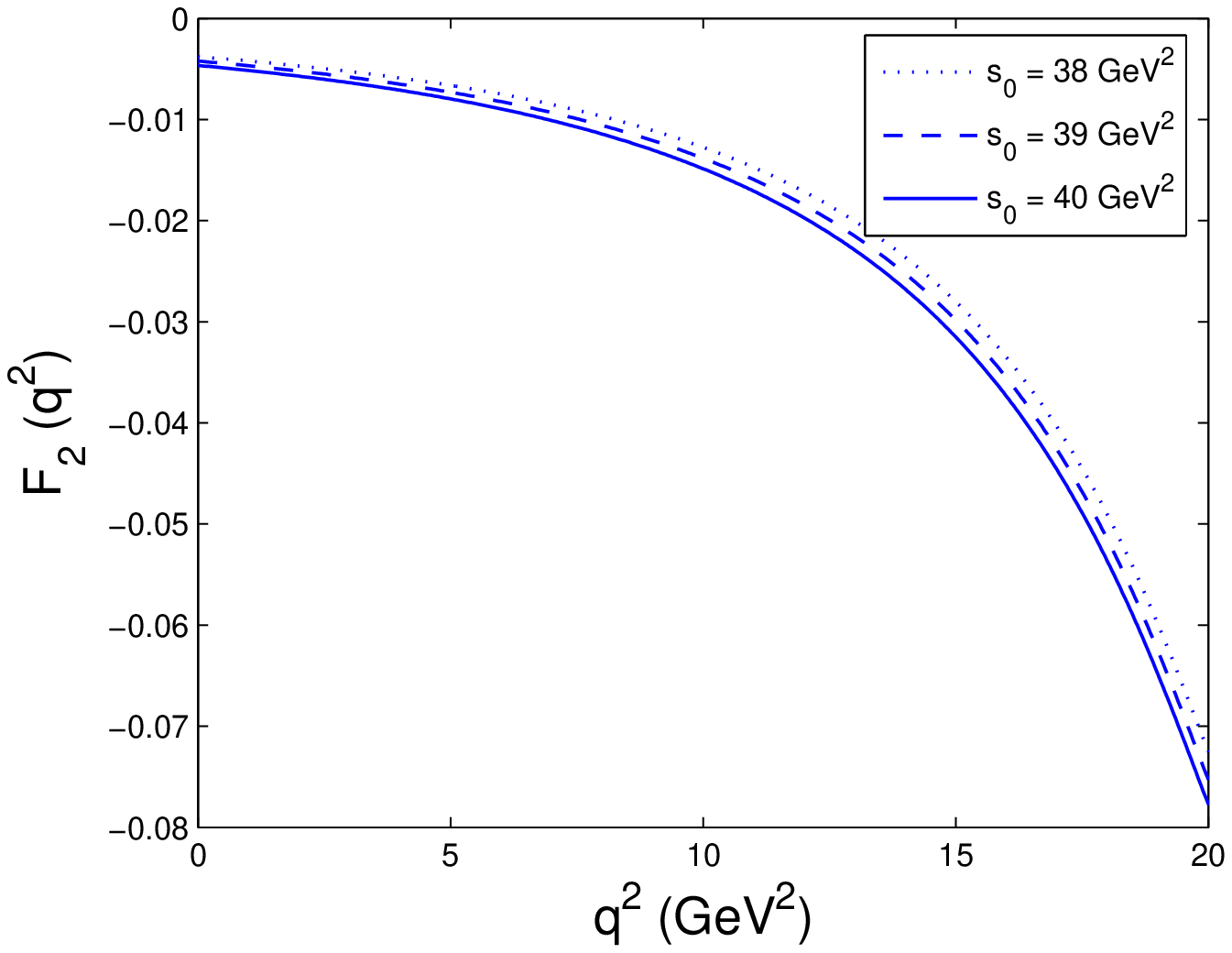}}
\end{minipage}
\begin{minipage}{7cm}
\epsfxsize=7cm \centerline{\epsffile{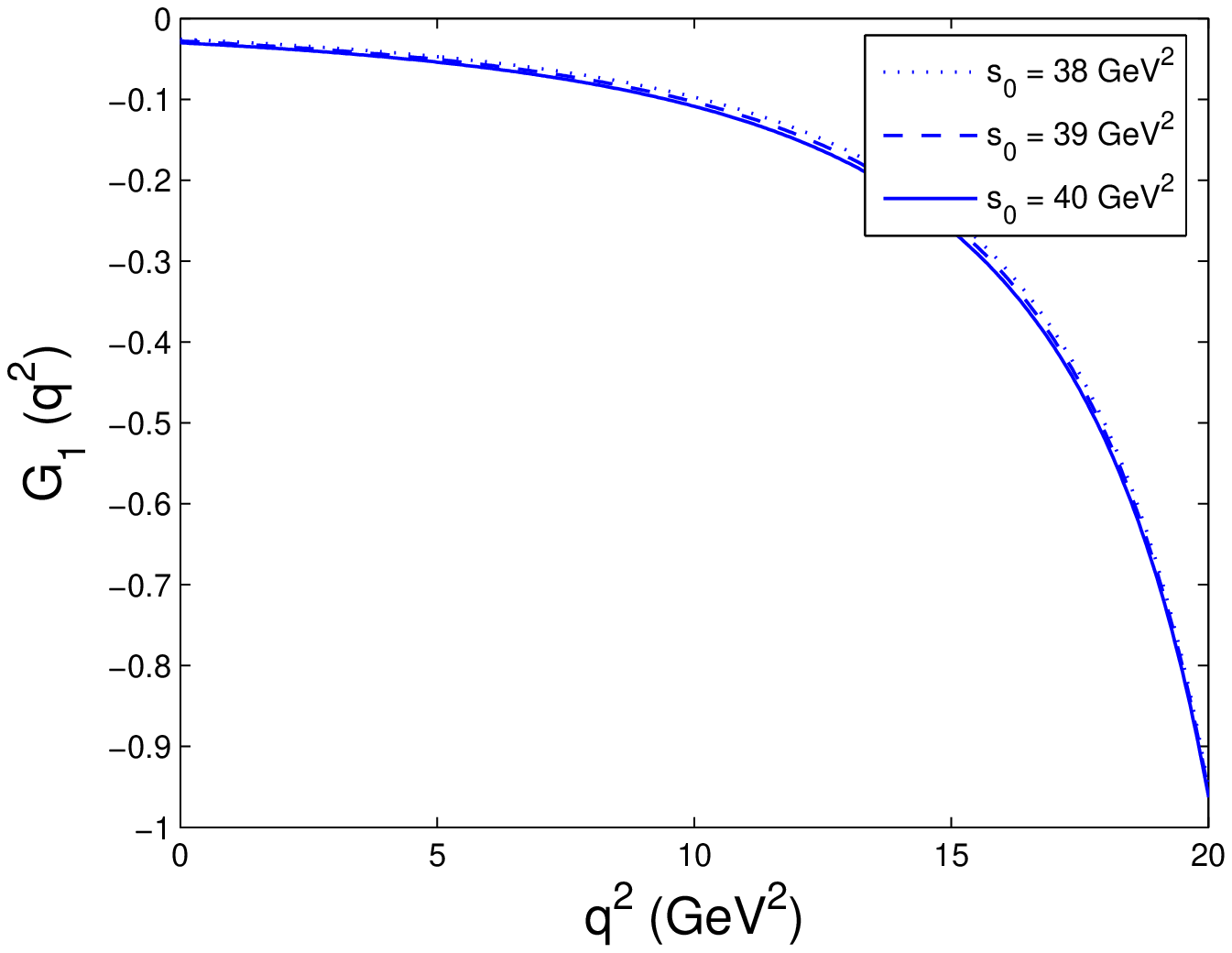}}
\end{minipage}
\begin{minipage}{7cm}
\epsfxsize=7cm \centerline{\epsffile{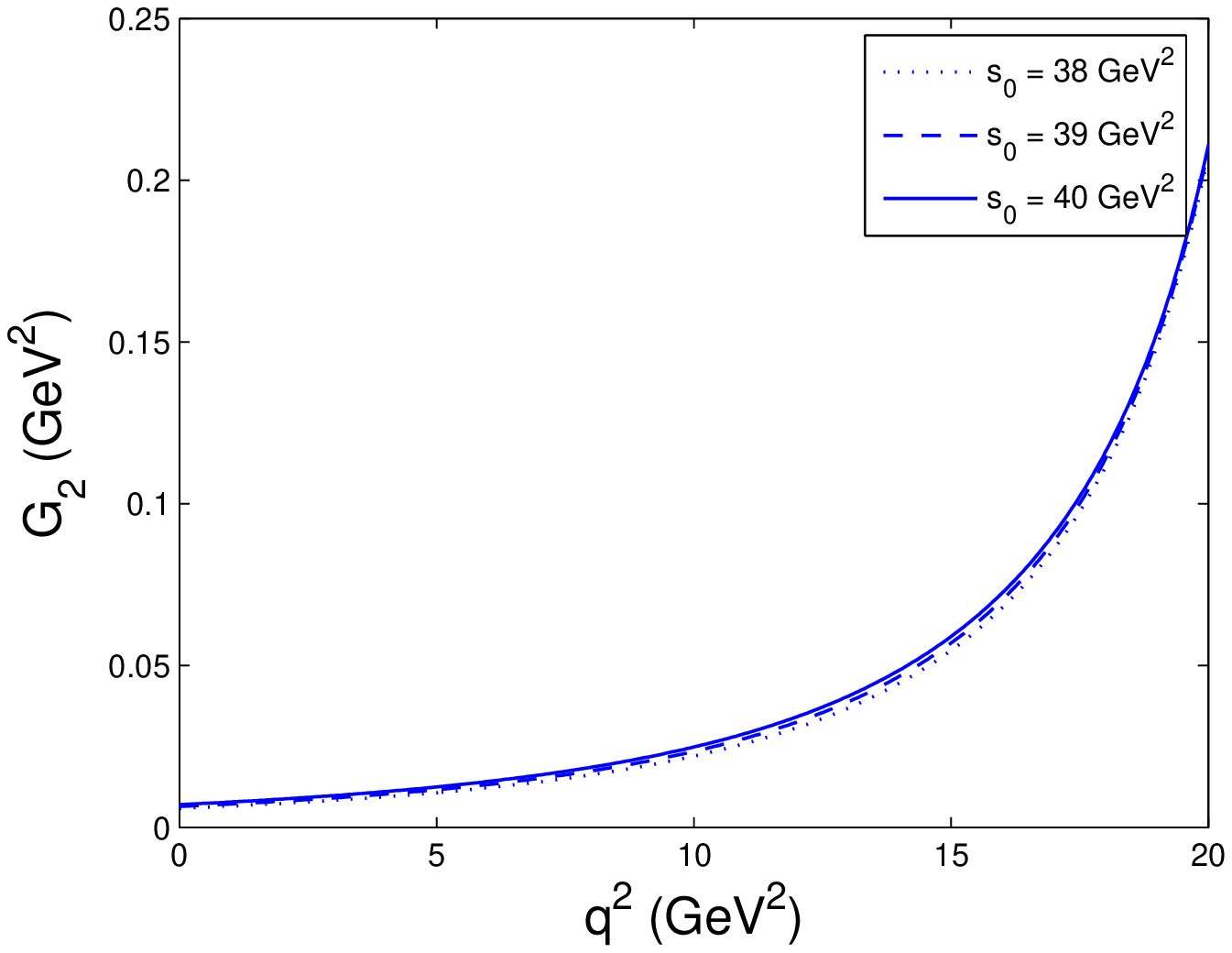}}
\end{minipage}
\caption{\quad Predictions of the form factors on the momentum transfer $q^2$ with $s_0=38,\,39,\,40\,\mbox{GeV}^2$ when Ioffe-type current is used. }\label{fig03}
\end{figure}

As we can see from Fig. \ref{fig01}, the dependence of the form factors on the Borel parameter $M_B^2$ is quite mild. It is noted that the windows of the Borel parameters vary along with the momentum transfer $q^2$. However, the windows are flat enough for us to choose a special value of the Borel parameter for the evaluation of the form factors. Hence, the Borel parameter is set to be $M_B^2=13\,\mbox{GeV}^2$ such that the higher resonance contributions are suppressed to be less than $30\%$. Keeping this constraint in mind, we calculate the $q^2$ dependence of all the form factors arising in the parametrization of the hadronic matrix elements based on the sum rules we derived in Sec. \ref{sec2}. The results are shown in Fig. \ref{fig02} where the CZ-type current for the $\Lambda_b$ baryon is adopted and Fig. \ref{fig03} where the Ioffe-type current is used.

As we know, the light-cone sum rules are only able to estimate the form factors within the region where they are reliable. Therefore we need to use the fitting formulas to extrapolate the curves to the whole dynamic region. Note that when the CZ-type current is used, the relationships: $g_1 = - f_1$, $g_2 = f_2$, $G_1 = - F_1$, and $G_2 = F_2$ hold. As has been done in literatures, we use the following double-pole formula to fit the resulting curves for these form factors:
\begin{equation}\label{fit01}
f_{i}(x) = \frac{f_{i}(0)}{1-a_1 q^2/M^2_{\Lambda_b}+a_2 q^4/M^2_{\Lambda_b}}\; ,
\end{equation}
where the parameters $f_{i}(0)$, $a_1$ and $a_2$ are given in the TABLE \ref{table01}. The system errors of the sum rule method are difficult to estimate so that they are not considered. Here we only give the errors that come from the choices of the thresholds and the sum rule windows.
\begin{table}
\caption{Parameters of the fitting formula for the form factors when the CZ-type current is used.}\label{table01}
\begin{tabular}{|c|c|c|c|c|c|}
  \hline
  $f_{i}$ & $f_{i}(0)$ & $a_1$ & $a_2$  \\
  \hline
  $f_1$ & $-0.0618^{+0.0015}_{-0.0041}$ & $2.93^{-0.0019}_{+0.0013}$ & $2.21^{-0.0016}_{+0.0013}$  \\
  $f_2$ & $0.172^{+0.013}_{-0.010}$ & $2.42^{-0.048}_{+0.055}$ & $1.55^{-0.046}_{+0.053}$  \\
  $F_1$ & $0.172^{+0.013}_{-0.010}$ & $2.42^{-0.048}_{+0.055}$ & $1.55^{-0.046}_{+0.053}$  \\
  $F_2$ & $-0.00675^{-0.00017}_{+0.00010}$ & $2.60^{-0.052}_{+0.075}$ & $1.76^{-0.058}_{+0.082}$  \\
  \hline
\end{tabular}
\end{table}
When the Ioffe-type current is used, the form factors can also be parameterized by the double-pole formula (\ref{fit01}), in which case all the parameters $f_{i}(0)$, $a_1$ and $a_2$ are given in the TABLE \ref{table02}.
\begin{table}
\caption{Parameters of the fitting formula for the form factors when the Ioffe-type current is used.}\label{table02}
\begin{tabular}{|c|c|c|c|c|c|}
  \hline
  $f_{i}$ & $f_{i}(0)$ & $a_1$ & $a_2$  \\
  \hline
  $f_1$ & $0.0454^{+0.0038}_{-0.0012}$ & $2.963^{-0.009}_{+0.002}$ & $2.258^{-0.010}_{+0.025}$ \\
  $f_2$ & $0.0348^{+0.0028}_{-0.0023}$ & $2.69^{-0.035}_{+0.037}$ & $1.84^{-0.039}_{+0.049}$  \\
  $g_1$ & $0.0464^{+0.0034}_{-0.0025}$ & $2.95^{-0.0018}_{+0.0045}$ & $2.24^{-0.0019}_{+0.0061}$ \\
  $g_2$ & $0.0301^{+0.0030}_{-0.0028}$ & $2.54^{-0.044}_{+0.052}$ & $1.70^{-0.048}_{+0.075}$  \\
  $F_1$ & $0.0355^{+0.0031}_{-0.0024}$ & $2.67^{-0.034}_{+0.024}$ & $1.85^{-0.035}_{+0.037}$  \\
  $F_2$ & $-0.00475^{-0.00059}_{+0.00061}$ & $2.628^{-0.041}_{+0.058}$ & $1.815^{-0.052}_{+0.078}$  \\
  $G_1$ & $-0.0346^{-0.0024}_{+0.0019}$ & $2.68^{-0.030}_{+0.039}$ & $1.83^{-0.036}_{+0.047}$  \\
  $G_2$ & $0.00790^{+0.00087}_{-0.00060}$ & $2.668^{-0.056}_{+0.043}$ & $1.815^{-0.074}_{+0.052}$  \\
  \hline
\end{tabular}
\end{table}

Using these fitting formulas for the form factors above, we can now calculate the decay widths of the processes $\Lambda_b\rightarrow\Lambda\gamma$ and $\Lambda_b\rightarrow\Lambda l^+l^-$. Let us first look at the process $\Lambda_b\rightarrow\Lambda\gamma$, which is related to the form factors $f_2$ and $g_2$ at the end point of zero momentum transfer. Averaging over the range $s_0=38-40\;\mbox{GeV}^2$ and ${M_B}^2=10-15\;\mbox{GeV}^{2}$, we get the value of the form factors as
\begin{equation}
f_2(0)=g_2(0)=0.160\pm 0.013\; ,
\end{equation}
for the CZ type interpolating current and
\begin{eqnarray}
f_2(0)= g_2(0)= 0.028\pm 0.003\; ,
\end{eqnarray}
for the Ioffe type interpolating current.

In order to calculate the decay width, we still need to know some other parameters. The Fermi coupling constant we used is $G_F = 1.166 \times 10^{-5}\;\mathrm{GeV}^2$ and the fine-structure constant is $1/137$. For the CKM matrix elements, we use the values given in Ref. \cite{PDG10} as $|V_{tb}|=0.9992$ and $|V_{ts}| = 0.0410$. And we use the Wilson coefficient $C_7(\mu=m_b)=-0.31$ in the SM \cite{wym}. The mass of the $s$ quark used in this paper is $m_s = 0.15\;\mathrm{GeV}$. As we have mentioned above, the relationships $g_V=1+m_s/m_b$ and $g_A=1-m_s/m_b$ hold in SM. Putting all these parameters into the formula (\ref{decayrate01}), we can obtain the decay widths in different cases as
\begin{eqnarray}
\Gamma_{\mathrm{CZ}}(\Lambda_b\rightarrow \Lambda\gamma)=(9.42^{+1.59}_{-1.47})\times
10^{-18}\;\mbox{GeV},\nonumber\\
\Gamma_{\mathrm{Ioffe}}(\Lambda_b\rightarrow \Lambda\gamma)=(2.88^{+0.66}_{-0.58})\times
10^{-19}\;\mbox{GeV}.
\end{eqnarray}
Considering that the life time of the $\Lambda_b$ baryon is about $1.391\times 10^{-12}\;\mathrm{s}$ \cite{PDG10}, the corresponding branching ratios can then be easily calculated as
\begin{eqnarray}
Br_{\mathrm{CZ}}(\Lambda_b\rightarrow\Lambda\gamma)=(1.99^{+0.34}_{-0.31})\times10^{-5},\nonumber\\
Br_{\mathrm{Ioffe}}(\Lambda_b\rightarrow\Lambda\gamma)=(0.61^{+0.14}_{-0.13})\times10^{-6}.
\end{eqnarray}
We have list our results in TABLE \ref{table03}, where results from other models are also presented. We can see that the choice of the interpolating current for the $\Lambda_b$ baryon brings in fundamental influence on the prediction for the branching ratio of the $\Lambda_b\rightarrow\Lambda\gamma$ decay. When the CZ-type current is adopted, the result is roughly in agreement with predictions from other methods. But it turns up to be much smaller when the Ioffe-type current is used. However, both the results are compatible with the upper limit $1.3\times 10^{-3}$ given by the Particle Data Group \cite{PDG10}.

\begin{table}[h]
\caption{Decay branching ratios ($B_r$) of $\Lambda_b\to \Lambda\gamma$ based on the various models}\label{table03}
 \centering
\begin{tabular}{|c|c|}
\hline
Model & $B_r$ \\
\hline
CZ current & $1.99^{+0.34}_{-0.31}\times 10^{-5}$\\
\hline
Ioffe current & $0.61^{+0.14}_{-0.13}\times 10^{-6}$\\
\hline
Pole Model\cite{Mannel}& $(0.10\sim 0.45)\times 10^{-5}$\\
\hline
QCD Sum Rule\cite{hcs} & $(3.7\pm 0.5)\times 10^{-5}$ \\
\hline
COQM\cite{Mohanta} & $0.23\times 10^{-5}$ \\
\hline
HQET\cite{HaiYang}& $(1.2\sim 1.9)\times 10^{-5}$ \\
\hline
bag model\cite{HaiYang}& $0.6\times10^{-5}$\\
\hline
PQCD\cite{lxq}& $(4.3\sim6.8)\times10^{-8}$\\
\hline
PDG\cite{PDG10}&$< 1.3\times 10^{-3}$\\
\hline
\end{tabular}
\end{table}

\begin{figure}
\begin{minipage}{7cm}
\epsfxsize=7cm \centerline{\epsffile{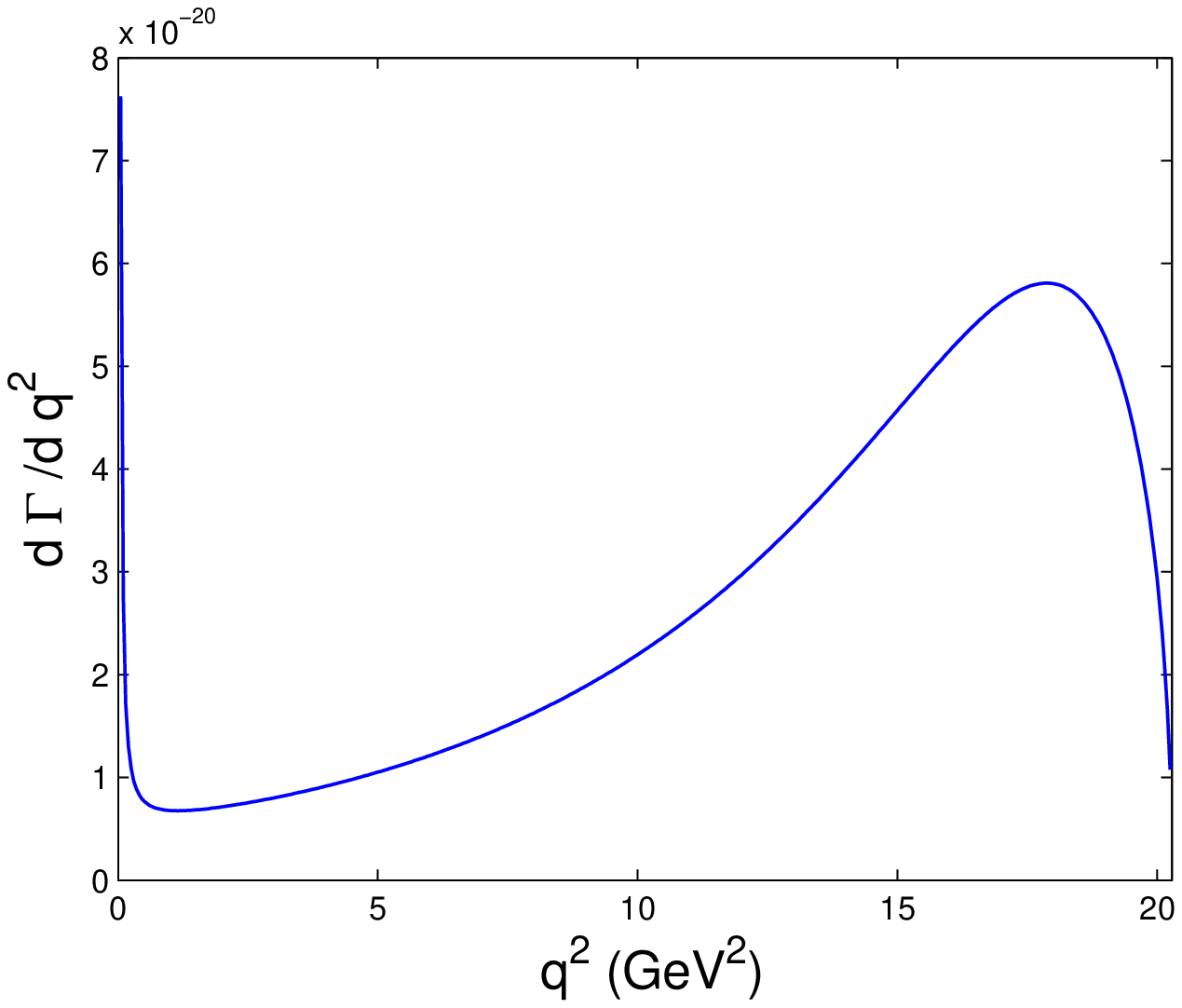}}
\end{minipage}
\begin{minipage}{7cm}
\epsfxsize=7cm \centerline{\epsffile{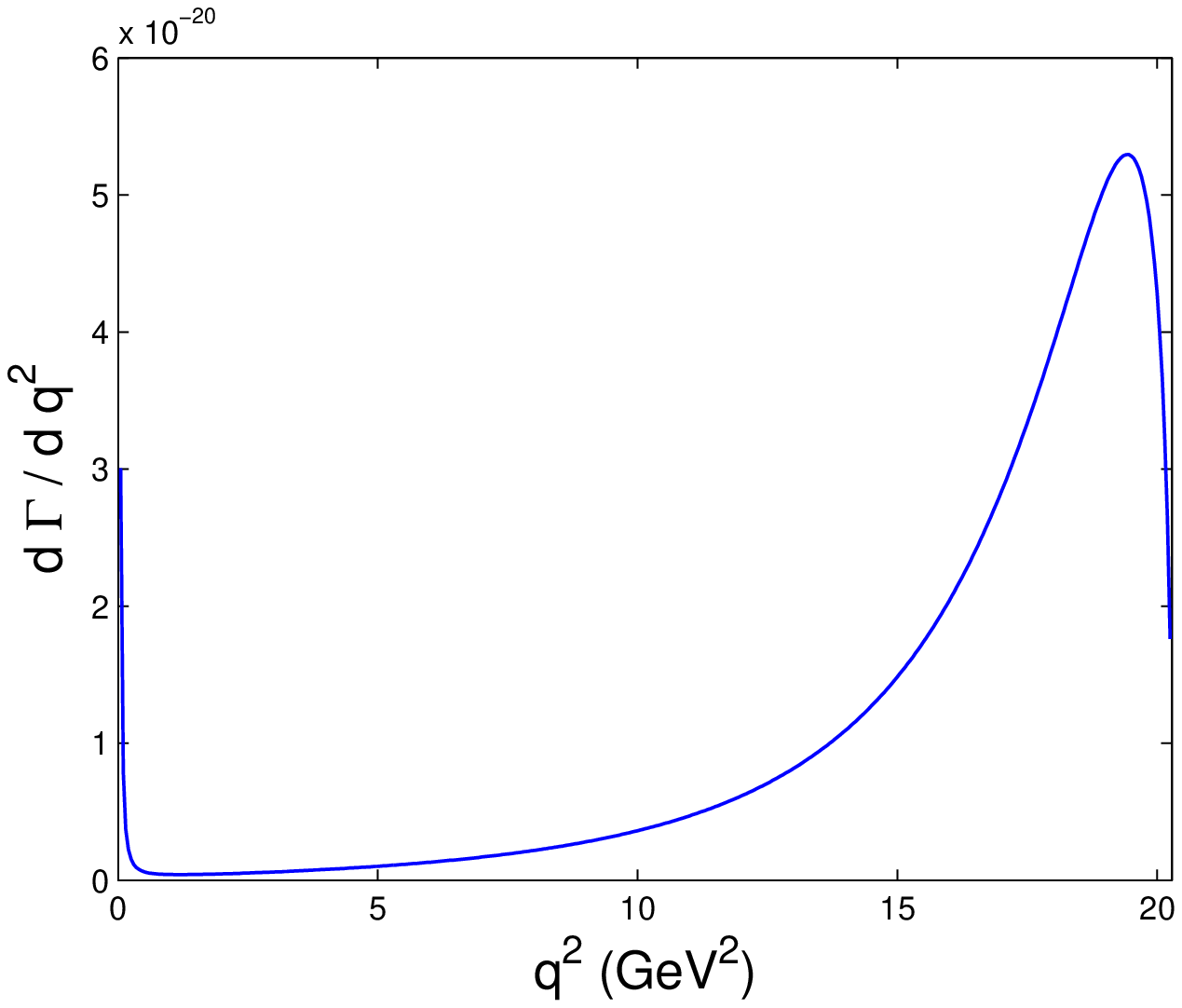}}
\end{minipage}
\caption{\quad Dependence of the differential decay width for the $\Lambda_b\rightarrow\Lambda l^+l^-$ decay on the momentum transfer $q^2$ in the CZ-type current case (on the left side) and the Ioffe-type current case (on the right side). }\label{fig04}
\end{figure}

Now let us proceed to evaluate the decay width and branching ratio of the $\Lambda_b\rightarrow\Lambda l^+l^-$ process. The Wilson coefficients in the effective Lagrangian are set to be $C_7^{\mathrm{eff}}(\mu=m_b) = -0.31$, $C_9^{\mathrm{eff}}(\mu=m_b) = 4.344$, and $C_{10}^{\mathrm{eff}}(\mu=m_b) = -4.669$ \cite{wym}. Put them and the fitting formulas of the form factors given above into Eq. (\ref{decayrate02}), we achieve the $q^2$ dependence of the differential decay rate in the whole dynamical region $0\,\mbox{GeV}^2<q^2<20.286\,\mbox{GeV}^2$, as plotted in Fig. \ref{fig04}. The appearance of a large enhancement at the end point $q^2=0$ is due to the the $1/q^2$ and $1/q^4$ factors in Eq. (\ref{decayrate02}). The $q^2$ distribution of the differential decay width in the CZ-type current case is consistent with that of Ref. \cite{wym} while in the Ioffe-type current case the distribution are comparatively more concentrated in the area of larger $q^2$.

The total decay width of this process can then be easily obtained by performing the integration over the dynamical region. The results are  $\Gamma(\Lambda_b\rightarrow\Lambda l^+l^-)_{\mathrm{CZ}}=1.87^{+0.18}_{-0.03}\times 10^{-18} \mathrm{GeV}$  for the CZ-type current and $\Gamma(\Lambda_b\rightarrow\Lambda l^+l^-)_{\mathrm{Ioffe}}=9.60^{+1.21}_{-0.41}\times 10^{-19} \mathrm{GeV}$ for the Ioffe type current, respectively. The corresponding branching ratios are displayed in TABLE \ref{table04}, where we have also shown the results from the literature. We can see from the table that the branching ratio in the CZ-type current case is totally consistent with the LCSR prediction given in Ref. \cite{Aliev} while the result in the Ioffe-type current case is in good agreement with the heavy quark effective theory (HQET) calculation \cite{Chen}. As there is no experimental data available, the rationality of the estimations can be understood from the comparison between the orders of magnitudes for the processes $\Lambda_b\rightarrow\Lambda l^+l^-$ and  $\Lambda_b\rightarrow \Lambda\gamma$. It is noted that we have neglected the mass of final leptons in our analysis, which should be good approximations for electrons and muons but not for tauons. In addition, we have adopted the usual quark-hadron duality ansatz for the hadronic spectral density in the derivation of the LCSRs, which is not accurate for the reason that contribution from the corresponding negative-parity resonance is not taken into account \cite{BCDN93,JKO96,KKMW11}. This will be considered in our next work.
\begin{table}[h]
\caption{Decay branching ratios ($B_r$) of $\Lambda_b\to \Lambda l^+l^-$  based on the various models}\label{table04}
 \centering
\begin{tabular}{|c|c|c|c|c|}
\hline
 & CZ current & Ioffe current & LCSR in Ref.\cite{Aliev}& HQET\cite{Chen} \\
\hline
Br & $3.96^{+0.38}_{-0.08}\times 10^{-6}$ & $2.03^{+0.26}_{-0.09}\times 10^{-6}$ & $(4.6\pm1.6)\times 10^{-6}$ & $(2.23\sim 3.34)\times 10^{-6}$\\
\hline
\end{tabular}
\end{table}

In summary, we have investigated the rare radiative decay $\Lambda_b\rightarrow\Lambda\gamma$ and rare semileptonic decay $\Lambda_b\rightarrow\Lambda l^+l^-$ of $\Lambda_b$ baryon within the framework of the standard model. Taking into account that two different types of interpolating currents for the $\Lambda_b$ baryon can be employed, we estimate the transition form factors in both cases using the light-cone sum rule approach. With these form factors, the decay widths and branching ratios are predicted. We find that the choice of the interpolating current for the $\Lambda_b$ baryon can affect the predictions significantly, especially for the rare radiative decay process. When the CZ-type current is employed, the decay widths are calculated to be $\Gamma(\Lambda_b\rightarrow\Lambda\gamma)=9.42^{+1.59}_{-1.47}\times 10^{-18} \mathrm{GeV}$ and $\Gamma(\Lambda_b\rightarrow\Lambda l^+l^-)=1.87^{+0.18}_{-0.03}\times 10^{-18} \mathrm{GeV}$. The branching ratio of $\Lambda_b\rightarrow\Lambda\gamma$ which we predict to be $B_r(\Lambda_b\rightarrow\Lambda\gamma)=1.99^{+0.34}_{-0.31}\times10^{-5}$ is in accord with the upper limit $1.3\times 10^{-3}$ given by the Particle Data Group while the branching ratio of the process $\Lambda_b\rightarrow\Lambda l^+l^-$ which we predict to be $B_r(\Lambda_b\rightarrow\Lambda l^+l^-)=3.96^{+0.38}_{-0.08}\times10^{-6}$ is in good agreement with the result from other LCSR calculations. When the Ioffe-type current is used, the decay widths turn up to be $\Gamma(\Lambda_b\rightarrow\Lambda\gamma)=2.88^{+0.66}_{-0.58}\times 10^{-19} \mathrm{GeV}$ and $\Gamma(\Lambda_b\rightarrow\Lambda l^+l^-)=9.60^{+1.21}_{-0.41}\times 10^{-19} \mathrm{GeV}$. The branching ratios of these weak processes are then predicted to be  $B_r(\Lambda_b\rightarrow\Lambda\gamma)=0.61^{+0.14}_{-0.13}\times10^{-6}$ which is also in accord with the above upper limit, and $B_r(\Lambda_b\rightarrow\Lambda l^+l^-)=2.03^{+0.26}_{-0.09}\times10^{-6}$ which is compatible with the HQET result.

\subsection*{Acknowledgements}
This work was supported in part by the National Natural Science
Foundation of China under Contract Nos. 11205242, 10975184, 11105222 and 11105223.
\appendix
\section{}\label{appendix01}
When the CZ-type current for the $\Lambda_b$ baryon is adopted, the light-cone sum rules for the form factors $f_1$, $f_2$, $F_1$, and $F_2$ are as follows:
\begin{eqnarray}
f_1f_{\Lambda_b} e^{-M_{\Lambda_b}^2/M_B^2}&=&
-\frac{Mq^2}{M_B^2}\int_{\alpha_{30}}^1d\alpha_3\frac{1}{\alpha_3}e^{-s/M_B^2}\Big\{B_1(\alpha_3)-
\frac{M^2}{M_B^2}B_3(\alpha_3)\Big\}\nonumber\\
&&-\frac{\alpha_{30}Mq^2e^{-s_0/M_B^2}}{m_b^2+
\alpha_{30}^2M^2-q^2}(B_1-\frac{M^2}{M_B^2}B_3)(\alpha_{30})\nonumber\\
&&-\frac{\alpha_{30}^2M^3q^2e^{-s_0/M_B^2}}{m_b^2+
\alpha_{30}^2M^2-q^2}[\frac{d}{d\alpha_{30}}\frac{\alpha_{30}}{m_b^2+
\alpha_{30}^2M^2-q^2}B_3({\alpha_{30}})],
\label{sumrule1}
\end{eqnarray}
\begin{eqnarray}
f_2f_{\Lambda_b} e^{-M_{\Lambda_b}^2/M_B^2}&=&
\int_{\alpha_{30}}^1d\alpha_3 e^{-s/M_B^2}\Big\{B_0-
\frac{M^2}{{M_B}^2}(B_1+B_2)(\alpha_3)+
\frac{M^4}{{M_B}^4}B_3(\alpha_3)\Big\}\nonumber\\
&&-\frac{\alpha_{30}^2M^2e^{-s_0/M_B^2}}{m_b^2+
\alpha_{30}^2M^2-q^2}(B_1+B_2-\frac{M^2}{M_B^2}B_3)(\alpha_{30})\nonumber\\
&&-\frac{\alpha_{30}^2M^4e^{-s_0/M_B^2}}{m_b^2+
\alpha_{30}^2M^2-q^2}\frac{d}{d\alpha_{30}}[\frac{\alpha_{30}^2}{m_b^2+
\alpha_{30}^2M^2-q^2}B_3(\alpha_{30})],
\label{sumrule2}
\end{eqnarray}
\begin{eqnarray}
F_1f_{\Lambda_b} e^{-M_{\Lambda_b}^2/M_B^2}&=&
\int_{\alpha_{30}}^1d\alpha_3 e^{-s/M_B^2}\Big\{B_0-
\frac{M^2}{{M_B}^2}(B_1+B_2)(\alpha_3)+
\frac{M^4}{{M_B}^4}B_3(\alpha_3)\Big\}\nonumber\\
&&-\frac{\alpha_{30}^2M^2e^{-s_0/M_B^2}}{m_b^2+
\alpha_{30}^2M^2-q^2}(B_1+B_2-\frac{M^2}{M_B^2}B_3)(\alpha_{30})\nonumber\\
&&-\frac{\alpha_{30}^2M^4e^{-s_0/M_B^2}}{m_b^2+
\alpha_{30}^2M^2-q^2}\frac{d}{d\alpha_{30}}[\frac{\alpha_{30}^2}{m_b^2+
\alpha_{30}^2M^2-q^2}B_3(\alpha_{30})],\nonumber\\
\label{sumrule3}
\end{eqnarray}
and
\begin{eqnarray}
F_2f_{\Lambda_b} e^{-M_{\Lambda_b}^2/M_B^2}&=&
\int_{\alpha_{30}}^1\frac{d\alpha_3}{\alpha_3}\frac{M}{M_B^2} e^{-s/M_B^2}\Big\{-
B_1(\alpha_3)+\frac{M^2}{{M_B}^2}B_3(\alpha_3)\Big\}\nonumber\\
&&+\frac{\alpha_{30}Me^{-s_0/M_B^2}}{m_b^2+
\alpha_{30}^2M^2-q^2}(-B_1+\frac{M^2}{M_B^2}B_3)(\alpha_{30})\nonumber\\
&&-\frac{\alpha_{30}^2M^3e^{-s_0/M_B^2}}{m_b^2+
\alpha_{30}^2M^2-q^2}\frac{d}{d\alpha_{30}}[\frac{\alpha_{30}}{m_b^2+
\alpha_{30}^2M^2-q^2}B_3(\alpha_{30})].\nonumber\\
\label{sumrule4}
\end{eqnarray}
In order to make the formulas clear and readable, we have used in the sum rules the short-hand notation $s=\frac{m_b^2}{\alpha_3}-\frac{1-\alpha_3}{\alpha_3}q^2+(1-\alpha_3)M_\Lambda^2$ and the following abbreviations
\begin{eqnarray}
B_0(\alpha_{3})&=&\int_0^{1-\alpha_3}d\alpha_1A_1(\alpha_1,1-\alpha_1-\alpha_3,\alpha_3),\nonumber\\
B_0'(\alpha_{3})&=&\int_0^{1-\alpha_3}d\alpha_1A_3(\alpha_1,1-\alpha_1-\alpha_3,\alpha_3),\nonumber\\
B_1(\alpha_{3})&=&-\widetilde A_1(\alpha_{3}) + \widetilde A_2(\alpha_{3}) - \widetilde A_3(\alpha_{3}),\nonumber\\
B_2(\alpha_{3})&=&-\widetilde A_1(\alpha_{3}) - \widetilde A_4(\alpha_{3}) + \widetilde A_5(\alpha_{3}),\nonumber\\
B_3(\alpha_{3})&=&\widetilde {\widetilde A_1}(\alpha_{3}) - \widetilde{\widetilde A_2}(\alpha_{3})+
\widetilde{\widetilde A_3}(\alpha_{3})+ \widetilde {\widetilde
A_4}(\alpha_{3})-\widetilde{\widetilde A_5}(\alpha_{3})+ \widetilde{\widetilde A_6}(\alpha_{3}),\nonumber\\
B_4(\alpha_{3})&=&-\widetilde A_1(\alpha_{3}) - \widetilde A_3(\alpha_{3}) + \widetilde A_5(\alpha_{3}),\nonumber\\
B_5(\alpha_{3})&=& \widetilde A_3(\alpha_{3}) - \widetilde A_4(\alpha_{3}).
\end{eqnarray}
The distribution amplitudes with tildes which come from the integration by parts in $\alpha_3$ are defined as
\begin{eqnarray}
\widetilde
A_i(\alpha_3)&=&\int_0^{\alpha_3}d{\alpha_3^\prime}\int_0^{1-\alpha_3^\prime}d\alpha_2
A_i(1-\alpha_2-\alpha_3^\prime,\alpha_2,\alpha_3^\prime), \nonumber\\
\widetilde{\widetilde
A_i}(\alpha_3)&=&\int_0^{\alpha_3}d{\alpha_3^\prime}\int_0^{\alpha_3^\prime}
d{\alpha_3^{\prime\prime}}
\int_0^{1-{\alpha_3^{\prime\prime}}}d\alpha_2A_i(1-\alpha_2-{\alpha_3^{\prime\prime}},
\alpha_2,{\alpha_3^{\prime\prime}}). \label{tilde}
\end{eqnarray}
We have used the partial integration in the variable $\alpha_3$ to eliminate the $1/(P\cdot x)$ factors appearing in the distribution amplitudes. With this being done, the surface terms sum up to zero.

In the sum rules (\ref{sumrule1})-(\ref{sumrule4}), $\alpha_{30}$ is connected with the continuum threshold $s_0$ via
\begin{equation}
\alpha_{30}=\frac{-(-q^2+s_0-{M_\Lambda}^2)+\sqrt{(-q^2+s_0-{M_\Lambda}^2)^2+
4(-q^2+m_b^2){M_\Lambda}^2}}{2{M_\Lambda}^2}.
\end{equation}

When the Ioffe-type current for the $\Lambda_b$ baryon is adopted, all the final sum rules for the form factors are as follows:
\begin{eqnarray}
f_1\lambda_{1b}M_{\Lambda_b}e^{-\frac{M_{\Lambda_b}^2}{M_B^2}}&&=\int_{\alpha_{30}}^1\frac{q^2d\alpha_3}{\alpha_3}e^{-\frac{s}{M_B^2}}
\Big\{B_0(\alpha_3)+\frac{M(\alpha_3M-m_b)}{\alpha_3M_B^2}B_1(\alpha_3)+\frac{m_bM^3}{\alpha_3M_B^4}B_3(\alpha_3)\}\nonumber\\
&&+\frac{Mq^2e^{-\frac{s_0}{M_B^2}}}{m_b^2+
\alpha_{30}^2M^2-q^2}\Big\{(\alpha_{30}M-m_b)B_1(\alpha_{30})+\frac{m_bM^2}{M_B^2}B_3(\alpha_{30})\Big\}\nonumber\\
&&-\frac{\alpha_{30}^2m_bM^3q^2e^{-\frac{s_0}{M_B^2}}}{m_b^2+\alpha_{30}^2M^2-q^2}\frac{d}{d\alpha_{30}}[\frac{1}{m_b^2+
\alpha_{30}^2M^2-q^2}B_3(\alpha_{30})],
\end{eqnarray}
\begin{eqnarray}
f_2\lambda_{1b}M_{\Lambda_b}e^{-\frac{M_{\Lambda_b}^2}{M_B^2}}&&=\int_{\alpha_{30}}^1\frac{d\alpha_3}{\alpha_3}e^{-\frac{s}{M_B^2}}
\Big\{m_bB_0(\alpha_3)+\frac{Mq^2}{\alpha_3M_B^2}B_1(\alpha_3)-\frac{m_bM^2}{M_B^2}(B_1+B_4+B_5)(\alpha_3)\nonumber\\
&&+\frac{m_bM^4}{M_B^4}B_3(\alpha_3)\}+\frac{e^{-\frac{s_0}{M_B^2}}}{m_b^2+
\alpha_{30}^2M^2-q^2}\Big\{Mq^2B_1(\alpha_{30})\nonumber\\
&&-\alpha_{30}m_bM^2(B_1+B_4+B_5)(\alpha_{30})+\frac{\alpha_{30}m_bM^4}{M_B^2}B_3(\alpha_{30})\Big\}\nonumber\\
&&-\frac{\alpha_{30}^2m_bM^4e^{-\frac{s_0}{M_B^2}}}{m_b^2+\alpha_{30}^2M^2-q^2}\frac{d}{d\alpha_{30}}[\frac{\alpha_{30}}{m_b^2+
\alpha_{30}^2M^2-q^2}B_3(\alpha_{30})],
\end{eqnarray}
\begin{eqnarray}
g_1\lambda_{1b}M_{\Lambda_b}e^{-\frac{M_{\Lambda_b}^2}{M_B^2}}&&=\int_{\alpha_{30}}^1\frac{q^2d\alpha_3}{\alpha_3}e^{-\frac{s}{M_B^2}}
\Big\{B_0(\alpha_3)+\frac{M(\alpha_3M-m_b)}{\alpha_3M_B^2}B_1(\alpha_3)-\frac{m_bM^3}{\alpha_3M_B^4}B_3(\alpha_3)\}\nonumber\\
&&+\frac{Mq^2e^{-\frac{s_0}{M_B^2}}}{m_b^2+
\alpha_{30}^2M^2-q^2}\Big\{(\alpha_{30}M-m_b)B_1(\alpha_{30})-\frac{m_bM^2}{M_B^2}B_3(\alpha_{30})\Big\}\nonumber\\
&&+\frac{\alpha_{30}^2m_bM^3q^2e^{-\frac{s_0}{M_B^2}}}{m_b^2+\alpha_{30}^2M^2-q^2}\frac{d}{d\alpha_{30}}[\frac{1}{m_b^2+
\alpha_{30}^2M^2-q^2}B_3(\alpha_{30})],
\end{eqnarray}
\begin{eqnarray}
g_2\lambda_{1b}M_{\Lambda_b}e^{-\frac{M_{\Lambda_b}^2}{M_B^2}}&&=\int_{\alpha_{30}}^1\frac{d\alpha_3}{\alpha_3}e^{-\frac{s}{M_B^2}}
\Big\{m_bB_0(\alpha_3)-\frac{Mq^2}{\alpha_3M_B^2}B_1(\alpha_3)-\frac{m_bM^2}{M_B^2}(B_1+B_4+B_5)(\alpha_3)\nonumber\\
&&-\frac{m_bM^4}{M_B^4}B_3(\alpha_3)\}+\frac{e^{-\frac{s_0}{M_B^2}}}{m_b^2+
\alpha_{30}^2M^2-q^2}\Big\{-Mq^2B_1(\alpha_{30})\nonumber\\
&&-\alpha_{30}m_bM^2(B_1+B_4+B_5)(\alpha_{30})-\frac{\alpha_{30}m_bM^4}{M_B^2}B_3(\alpha_{30})\Big\}\nonumber\\
&&+\frac{\alpha_{30}^2m_bM^4e^{-\frac{s_0}{M_B^2}}}{m_b^2+\alpha_{30}^2M^2-q^2}\frac{d}{d\alpha_{30}}[\frac{\alpha_{30}}{m_b^2+
\alpha_{30}^2M^2-q^2}B_3(\alpha_{30})],
\end{eqnarray}
\begin{eqnarray}
F_1\lambda_{1b}M_{\Lambda_b}e^{-\frac{M_{\Lambda_b}^2}{M_B^2}}&&=\int_{\alpha_{30}}^1\frac{d\alpha_3}{\alpha_3}e^{-\frac{s}{M_B^2}}
\Big\{(m_bB_0+MB_1+\alpha_3MB_0')(\alpha_3)\nonumber\\
&&-\frac{1}{\alpha_3M_B^2}(Mq^2B_1+2\alpha_3m_bM^2B_2+\alpha_3^2M^3B_5)(\alpha_3)+\frac{m_bM^4}{M_B^4}B_3(\alpha_3)\}\nonumber\\
&&+\frac{e^{-\frac{s_0}{M_B^2}}}{m_b^2+
\alpha_{30}^2M^2-q^2}\Big\{(Mq^2B_1+2\alpha_{30}m_bM^2B_2+\alpha_{30}M^3B_5)(\alpha_{30})\nonumber\\
&&+\frac{\alpha_{30}m_bM^4}{M_B^2}B_3(\alpha_{30})\Big\}
-\frac{\alpha_{30}^2m_bM^4e^{-\frac{s_0}{M_B^2}}}{m_b^2+\alpha_{30}^2M^2-q^2}
\frac{d}{d\alpha_{30}}[\frac{\alpha_{30}}{m_b^2+
\alpha_{30}^2M^2-q^2}B_3(\alpha_{30})],\nonumber\\
\end{eqnarray}
\begin{eqnarray}
F_2\lambda_{1b}M_{\Lambda_b}e^{-\frac{M_{\Lambda_b}^2}{M_B^2}}&&=\int_{\alpha_{30}}^1\frac{d\alpha_3}{\alpha_3}e^{-\frac{s}{M_B^2}}
\Big\{-B_0(\alpha_3)+\frac{M}{\alpha_3M_B^2}(\alpha_3M(B_1+B_5)+m_bB_1)(\alpha_3)\nonumber\\
&&+\frac{m_bM^3}{\alpha_3M_B^4}B_3(\alpha_3)\}+\frac{e^{-\frac{s_0}{M_B^2}}}{m_b^2+\alpha_{30}^2M^2-q^2}\Big\{\alpha_{30}M^2(B_1+B_5)(\alpha_{30})\nonumber\\
&&+m_bMB_1(\alpha_{30})+\frac{m_bM^3}{M_B^2}B_3(\alpha_{30})\Big\}\nonumber\\
&&-\frac{\alpha_{30}^2m_bM^3e^{-\frac{s_0}{M_B^2}}}{m_b^2+\alpha_{30}^2M^2-q^2}\frac{d}{d\alpha_{30}}[\frac{1}{m_b^2+
\alpha_{30}^2M^2-q^2}B_3(\alpha_{30})],
\end{eqnarray}
\begin{eqnarray}
G_1\lambda_{1b}M_{\Lambda_b}e^{-\frac{M_{\Lambda_b}^2}{M_B^2}}&&=-\int_{\alpha_{30}}^1\frac{d\alpha_3}{\alpha_3}e^{-\frac{s}{M_B^2}}
\Big\{(m_bB_0-MB_1-\alpha_3MB_0')(\alpha_3)\nonumber\\
&&+\frac{1}{\alpha_3M_B^2}(Mq^2B_1-2\alpha_3m_bM^2B_2+\alpha_3^2M^3B_5)(\alpha_3)+\frac{m_bM^4}{M_B^4}B_3(\alpha_3)\}\nonumber\\
&&-\frac{e^{-\frac{s_0}{M_B^2}}}{m_b^2+
\alpha_{30}^2M^2-q^2}\Big\{(Mq^2B_1-2\alpha_{30}m_bM^2B_2+\alpha_{30}^2M^3B_5)(\alpha_{30})\nonumber\\
&&+\frac{\alpha_{30}m_bM^4}{M_B^2}B_3(\alpha_{30})\Big\}+\frac{\alpha_{30}^2m_bM^4e^{-\frac{s_0}{M_B^2}}}{m_b^2
+\alpha_{30}^2M^2-q^2}\frac{d}{d\alpha_{30}}[\frac{\alpha_{30}}{m_b^2+
\alpha_{30}^2M^2-q^2}B_3(\alpha_{30})],\nonumber\\
\end{eqnarray}
and
\begin{eqnarray}
G_2\lambda_{1b}M_{\Lambda_b}e^{-\frac{M_{\Lambda_b}^2}{M_B^2}}&&=\int_{\alpha_{30}}^1\frac{d\alpha_3}{\alpha_3}e^{-\frac{s}{M_B^2}}
\Big\{B_0(\alpha_3)-\frac{M}{\alpha_3M_B^2}(\alpha_3M(B_1-B_5)-m_bB_1)(\alpha_3)\nonumber\\
&&+\frac{m_bM^3}{\alpha_3M_B^4}B_3(\alpha_3)\}-\frac{e^{-\frac{s_0}{M_B^2}}}{m_b^2+\alpha_{30}^2M^2-q^2}\Big\{\alpha_{30}M^2(B_1-B_5)(\alpha_{30})\nonumber\\
&&-m_bMB_1(\alpha_{30})-\frac{m_bM^3}{M_B^2}B_3(\alpha_{30})\Big\}\nonumber\\
&&-\frac{\alpha_{30}^2m_bM^3e^{-\frac{s_0}{M_B^2}}}{m_b^2+\alpha_{30}^2M^2-q^2}\frac{d}{d\alpha_{30}}[\frac{1}{m_b^2+
\alpha_{30}^2M^2-q^2}B_3(\alpha_{30})].
\end{eqnarray}


\end{document}